\documentclass[11pt]{article}
\usepackage[a4paper,margin=1in]{geometry}
\usepackage{times}
\usepackage{amsmath,amssymb}
\usepackage[authoryear,round]{natbib}
\usepackage{xcolor}
\usepackage{amsmath,amsfonts,bm}

\def\eqref#1{equation~\ref{#1}}
\def\1{\bm{1}}

\DeclareMathAlphabet{\mathsfit}{\encodingdefault}{\sfdefault}{m}{sl}
\SetMathAlphabet{\mathsfit}{bold}{\encodingdefault}{\sfdefault}{bx}{n}

\usepackage{graphicx}
\usepackage{hyperref}
\usepackage{url}
\usepackage{enumitem}
\usepackage{multirow}
\usepackage{wrapfig}
\usepackage{float}
\usepackage{subcaption}
\usepackage{rotating}

\title{ChoreoStudio: Planning What, When, and Who in Music-Driven Group Choreography}
\author{Jing Xu \quad Cunjian Chen\thanks{Corresponding author: \texttt{cunjian.chen@monash.edu}} \quad Qiuhong Ke\\[0.5em]
Department of Data Science and Artificial Intelligence, Faculty of Information Technology,\\
Monash University, Australia}
\date{}

\begin{document}
\typeout{TEXTFLOATSEP = \the\textfloatsep}
\maketitle

\begin{abstract}
Music-driven group dance generation must decide not only how each dancer moves, but also \emph{what} formation the group should form, \emph{when} it should change, and \emph{who} should execute each transition. Existing methods can generate orderly and rhythmically synchronized group motion, but remain limited in formation evolution and inter-dancer interaction. In human choreography, formation transitions often align with changes in high-level musical structure, such as the progression from an introduction to a chorus or from a chorus to a climax. We present \textbf{ChoreoStudio}, which makes choreography planning explicit. A music disentangler first decomposes the audio into frame-level rhythmic features and segment-level structural features. Given the segment-level features and sparse labels, a frozen omni-modal model acts as a scriptwriter, producing a global semantic description that guides the Director's choreography planning. Based on this context, the Director produces three coupled interaction decisions: a centroid-relative point set specifies \emph{what} formation to produce, a transition gate determines \emph{when} it should change, and a mobility allocation indicates \emph{who} should execute the transition. The transition gate and mobility allocation are further conditioned on the planned geometric change, coupling formation targets with their temporal and dancer-specific execution. At a finer temporal scale, frame-aligned music features condition the diffusion generator to control detailed body motion and preserve beat-level alignment. The segment-level interaction plan is injected in parallel through frame-aligned instruction tokens and cross-dancer attention biases, coordinating how the generator realizes each transition. Experiments on AIOZ-GDance demonstrate that ChoreoStudio follows the planned interactions while achieving state-of-the-art performance in motion quality, group organization, and formation dynamics.
\end{abstract}

\vspace{-3pt}
\section{Introduction}
\label{sec:introduction}

Music-driven group dance generation aims to synthesize coordinated 3D motions for multiple dancers from music. Unlike single-dancer generation, this task must determine not only how each dancer moves, but also how their interactions form an intentional cohesive group performance. Effective choreography assigns dancers distinct spatial roles, organizes their formation over time, and determines when and by whom each transition is executed. Individually plausible and synchronized movement is therefore insufficient on its own to create a cohesive group dance.

Existing group-dance methods coordinate dancers through joint generation~\citep{le2023music,le2023controllable}, coherence-oriented training objectives~\citep{yang2024codancers}, trajectory conditioning~\citep{dai2025harmonious,dai2026tcdiffpp}, and spatial-temporal interaction modeling~\citep{xu2025stgdance,xu2026st}. Across these designs, coordination is typically induced within the generator by jointly processing latent pose, trajectory, and interaction features across dancers. Although these mechanisms improve local consistency and group coherence, they leave two key choreography decisions unresolved. First, dancers are exchangeable before generation, whereas choreography assigns asymmetric roles such as leader, flank, or rear dancer. Symmetric cross-dancer attention does not determine which dancer should assume each role, while initializing dancers with independent noise breaks this symmetry arbitrarily rather than through deliberate role assignment. Second, inter-dancer spatial formation is not represented as an explicit object to be composed and assigned. It remains implicit in latent interactions or is replaced by externally specified trajectories, leaving the model without a direct representation of the intended group arrangement.

We expose the consequence of this entanglement by analyzing the temporal spectrum of inter-dancer geometry. Decomposing pairwise ground-plane distances into slow formation dynamics and fast residual motion shows that existing generators retain substantial high-frequency activity but markedly underrepresent the low-frequency evolution present in real dances (Appendix~\ref{app:formation_analysis}). In other words, they can keep dancers moving and locally coordinated without producing sufficiently rich macro-level formation change. This suggests that the key missing capability is not simply stronger cross-dancer interaction or smoother motion, but an \emph{explicit choreography abstraction} between music and motion: a representation that specifies what the group should become, when it should change, and who should execute that change, before deciding how the corresponding full-body motion should look.

\begin{figure}[t]
    \centering
    \includegraphics[width=0.92\linewidth]{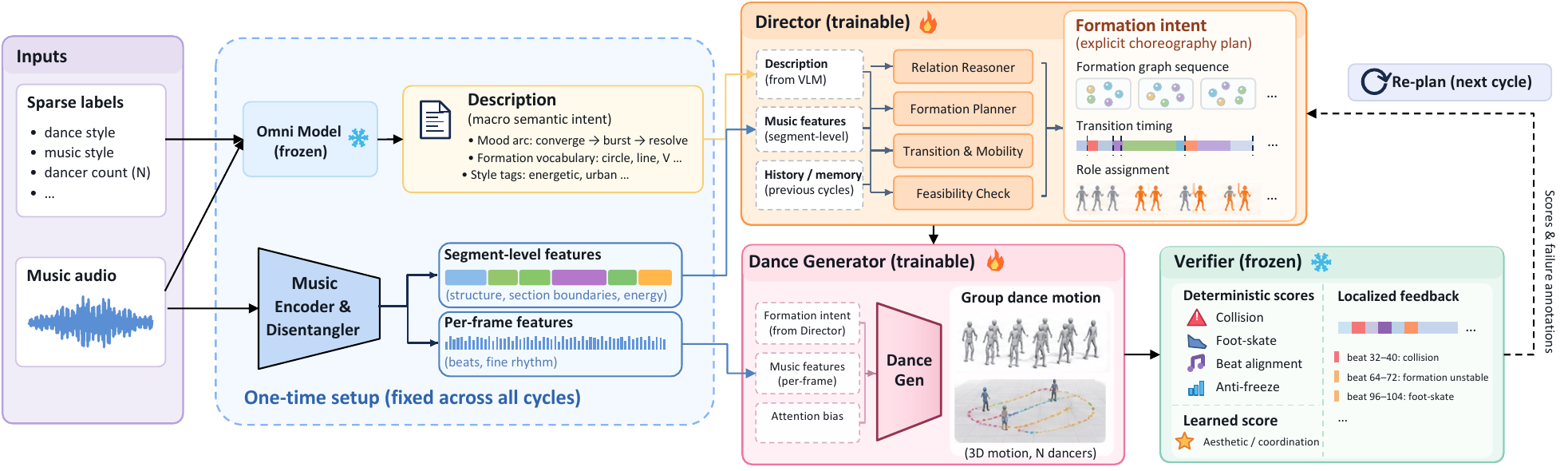}
    \caption{\textbf{Overview of ChoreoStudio.} Multi-scale music features and a frozen omni-modal model provide rhythmic, structural, and semantic context. The Director jointly constructs a directed inter-dancer relation graph and a centroid-relative formation point set, then uses the planned geometric change to schedule transition timing and dancer mobility. The resulting plan conditions the Dance Generator through frame-aligned dancer-specific instructions and relation-derived spatial attention biases.}
    \label{fig:workflow}
\end{figure}

Motivated by this observation, we introduce \textbf{ChoreoStudio}, which separates choreography composition from motion realization through an explicit, executable plan (Figure~\ref{fig:workflow}). The design follows a simple principle: decide \emph{what--when--who} at the choreography level, then synthesize \emph{how} at the motion level. To support these different temporal scales, a learned music disentangler extracts segment-level structural features for planning and frame-level rhythmic features for detailed synthesis; a frozen audio-language model additionally supplies high-level semantic intent. A group-dance Director converts this context into a structured plan containing relative formation geometry, directed dancer relations, transition timing, spatial roles, and mobility allocation. Importantly, these quantities are coupled: the magnitude of the intended geometric change informs whether a transition should occur, and the resulting scheduled displacement informs which dancers should carry it.

To ensure that these decisions are realized in full-body motion,
ChoreoStudio uses an explicit plan-to-motion interface. It converts the planned
formation, transition timing, and dancer mobility into dancer-specific guidance,
so that exchangeable dancers do not need to rediscover their roles and group
arrangement during denoising. No unrestricted Director state bypasses the plan,
making it the sole choreography-control path to the generator. The plan remains
fixed during denoising and can be inspected or edited at inference under fixed
music and sampling noise.

We evaluate ChoreoStudio on AIOZ-GDance under a shared 150-frame protocol, with long-sequence results reported separately. Beyond standard motion and group metrics, we measure formation dynamics, plan realization, and the causal response to controlled plan interventions. The results show improved motion quality and group organization over strong baselines, while frequency-based analysis confirms that these gains stem from realistic macro-level formation evolution rather than high-frequency distance fluctuations.

Our contributions are threefold:
\begin{itemize}[
    leftmargin=1.2em,
    labelsep=0.4em,
    topsep=0pt,
    itemsep=1pt,
    parsep=0pt,
    partopsep=0pt
]
    \item \textbf{We reveal a temporal-scale failure in group dance generation.} Frequency decomposition of inter-dancer geometry shows that existing generators retain {local motion activity} while underrepresenting macro-level formation evolution, exposing a gap unmeasured by conventional collision and correlation metrics.

    \item \textbf{We introduce unified choreographic planning for formation, timing, and mobility.} Our group-dance Director couples a directed inter-dancer relation graph with a geometrically valid formation point set, then conditions transition timing and dancer mobility on the planned spatial transformation. All plan targets are extracted from motion without manual choreography labels.

    \item \textbf{We make explicit choreography executable and testable.} Frame-aligned dancer instructions and relation-derived attention biases connect the structured plan to full-body generation without an unrestricted hidden-state bypass. Quantitative comparisons and controlled interventions demonstrate improved group organization and verify that generated motion responds to planned formation and timing changes.

\end{itemize}

\vspace{-7pt}
\section{Related Work}
\label{sec:related}

\vspace{-6pt}
\subsection{Music-Driven Single and Group Dance Generation}

\vspace{-4pt}
Single-person methods model music-motion correspondence through cross-modal
transformers such as FACT~\citep{li2021ai}, discrete representations such as
Bailando~\citep{siyao2022bailando}, and editable diffusion such as
EDGE~\citep{tseng2023edge}; DuetGen~\citep{ghosh2025duetgen} extends generation
to interacting pairs. For larger groups, GDanceR~\citep{le2023music},
GCD~\citep{le2023controllable}, and CoDancers~\citep{yang2024codancers} improve
coordination through joint, contrastive, and coherence-oriented modeling.
TCDiff variants~\citep{dai2025harmonious,dai2026tcdiffpp} introduce trajectory
and footwork control, while ST-GDance
variants~\citep{xu2025stgdance,xu2026st} support scalable long-duration
generation. However, formation evolution remains implicit or reduced to
low-level trajectories. ChoreoStudio instead jointly plans formation, transition
timing, and dancer mobility.

\vspace{-6pt}
\subsection{Multi-Scale Music Conditioning}

\vspace{-4pt}
Music-driven dance models typically use frame-aligned acoustic features for
local rhythmic synchronization~\citep{li2021ai,siyao2022bailando,tseng2023edge},
but provide limited guidance over long-term choreography. Hierarchical methods
incorporate global context or choreographic primitives
~\citep{li2025lodgepp,li2025melodance}, yet mainly organize individual motion.
ChoreoStudio instead disentangles music into segment-level and frame-level
conditions. A frozen omni-modal language model~\citep{Qwen3-Omni} interprets
music segments and sparse labels to guide formation, transition, and mobility
planning, while frame-level features preserve detailed beat alignment.

\vspace{-6pt}
\subsection{Structured Planning for Generative Motion}

\vspace{-4pt}
Recent motion systems increasingly separate semantic reasoning from motion synthesis. Language-centered models unify motion understanding, generation, and editing through LLM interfaces~\citep{jiang2023motiongpt,zhou2024avatargpt,wang2026motiongpt,wu2025motion,wu2025mg,zhu2026motiongpt3}. Structured-control methods organize individual motion using temporal tracks, atomic action schedules, or anchored motion phrases~\citep{petrovich2024multi,cai2026music,tang2026customdance}. NEWTON~\citep{feng2026newton} further demonstrates planner-generator-verifier orchestration for video synthesis. These methods plan individual motion or generation procedures, whereas ChoreoStudio jointly plans group formation geometry, transition timing, and dancer mobility for coordinated full-body motion.

\vspace{-6pt}
\section{Method}
\label{sec:method}

\vspace{-4pt}
\subsection{Overview and Problem Formulation}
\label{sec:overview}

\vspace{-4pt}
Given a music sequence and target group size, our goal is to generate a coordinated group dance:
\begin{equation}
\mathbf{X}
=
\{\mathbf{x}_{i,t}\}
\in
\mathbb{R}^{N \times L \times D_x},
\end{equation}
where $N$, $L$, and $D_x$ denote the numbers of dancers, motion frames, and
motion features, respectively, and $\mathbf{x}_{i,t}$ contains the root
translation, joint rotations, and foot contacts of dancer $i$ at frame $t$. Rather than mapping music directly to motion, ChoreoStudio introduces an explicit, executable choreography plan between musical interpretation and full-body motion synthesis.
%ChoreoStudio inserts explicit choreographic planning between musica interpretation and full-body motion realization.

\begin{figure*}[t]
    \centering
    \includegraphics[width=0.87\textwidth]{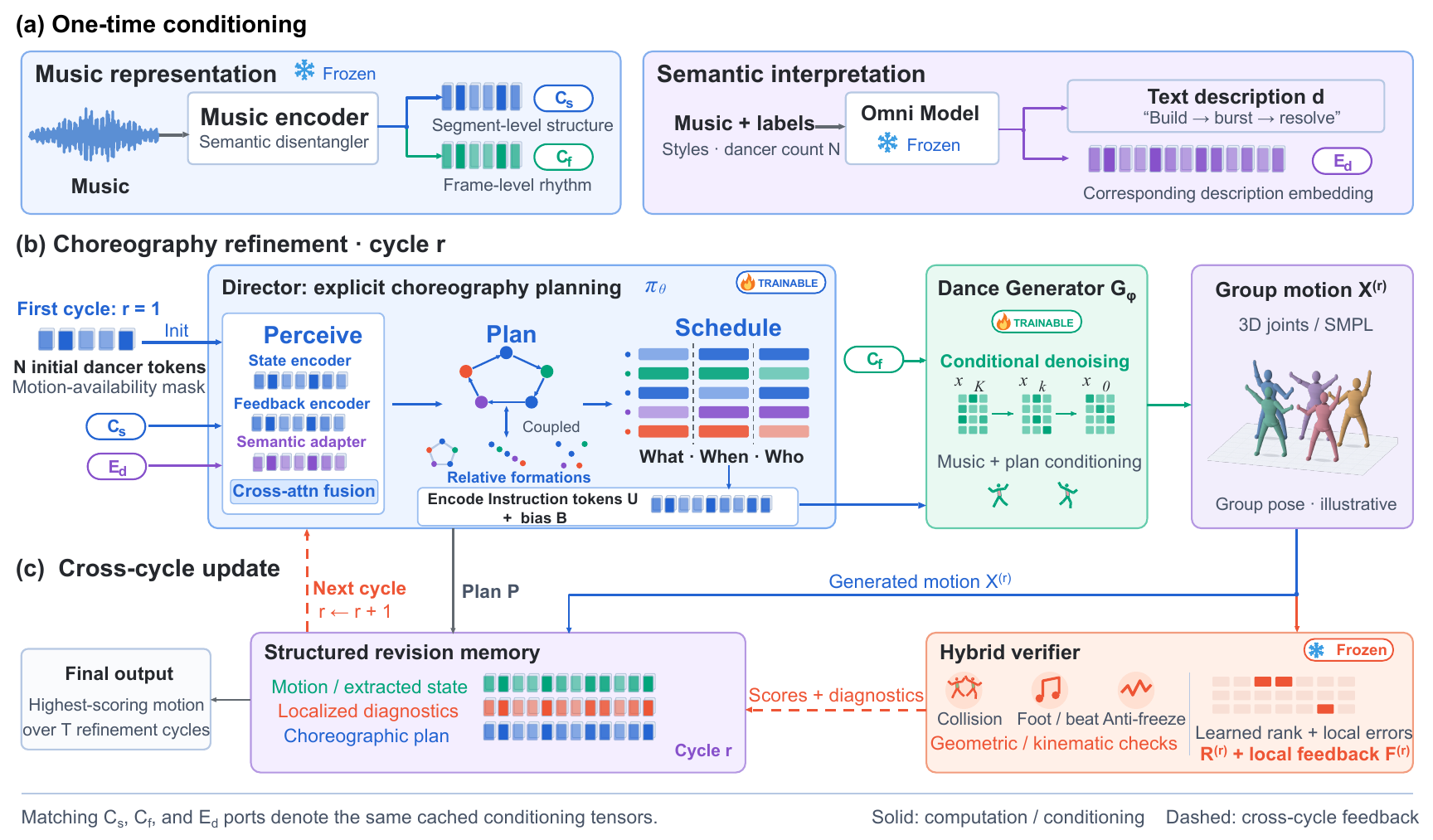}
    \caption{\textbf{ChoreoStudio architecture.} Segment-level structure and semantics guide the Director, while frame-level rhythm conditions motion synthesis; a mask marks unavailable motion. Through Perceive, Plan, Schedule, and Encode, the Director couples formation, timing, and mobility, encoding $\mathbf{P}$ as frame-aligned instructions $\mathbf{U}$ and relation-derived cross-dancer attention bias $\mathbf{B}$.}
    \label{fig:framework}
\end{figure*}

%As shown in Figure~\ref{fig:framework}, 
As shown in Figure~\ref{fig:framework}, ChoreoStudio operates in a closed planning--generation--verification loop. A music encoder followed by a trainable music disentangler extracts frame-level rhythmic features $\mathbf{C}_{f}$ for motion realization and segment-level structural features $\mathbf{C}_{s}$ for choreography planning. In parallel, the frozen omni-modal model Qwen3-Omni~\citep{Qwen3-Omni} converts the music audio and sparse attributes into a high-level choreography description $d$, which is embedded as semantic tokens $\mathbf{E}_{d}$. At refinement cycle $r$, the Director conditions on $\mathbf{C}_{s}$, $\mathbf{E}_{d}$, the current group state $\mathcal{S}^{(r)}$, and feedback from the preceding cycle to predict an explicit plan.
\begin{equation}
\mathbf{P}^{(r)}
=
\left\{
\mathbf{Y}^{(r)},
\mathbf{A}^{(r)},
\boldsymbol{\rho}^{(r)},
\mathbf{g}^{(r)},
\boldsymbol{\mu}^{(r)},
\boldsymbol{\delta}^{(r)}
\right\},
\end{equation}
where $\mathbf{Y}^{(r)}$ and $\mathbf{A}^{(r)}$ represent relative formations
and directed dancer relations, $\boldsymbol{\rho}^{(r)}$ encodes spatial roles,
$\mathbf{g}^{(r)}$ schedules transitions, $\boldsymbol{\mu}^{(r)}$ allocates
dancer mobility, and $\boldsymbol{\delta}^{(r)}$ controls pose divergence.
The plan is encoded into frame-aligned dancer instructions $\mathbf{U}^{(r)}$ and relation-aware attention biases $\mathbf{B}^{(r)}$, which condition the Dance Generator to realize $\mathbf{P}^{(r)}$ as full-body motion $\mathbf{X}^{(r)}$. A frozen Verifier then checks the generated motion for plan consistency, physical plausibility, and music alignment. Localized failures are then fed back to the Director to revise the next plan. Thus, each cycle explicitly separates deciding \emph{what} the group should do from synthesizing \emph{how} the dancers execute it. We omit cycle superscripts in single-cycle derivations.

\vspace{-6pt}
\subsection{Multi-Scale Music Conditioning and Semantic Intent}
\label{sec:conditioning}

\vspace{-4pt}
\textbf{Learned multi-scale music representations.}
Let $\mathbf{M}\in\mathbb{R}^{L_m\times D_m}$ denote acoustic features
pre-extracted with Librosa~\citep{mcfee2015librosa}. The music encoder
$E_{\mathrm{mus}}$ produces a frame-aligned representation:
\begin{equation}
\mathbf{Z}
=
E_{\mathrm{mus}}(\mathbf{M})
\in
\mathbb{R}^{L\times D}.
\end{equation}
The trainable disentangler $D_{\omega}$ projects $\mathbf{Z}$ into frame-level rhythmic and pooled segment-level structural representations:
\begin{equation}
\left(\mathbf{C}_{f},\mathbf{C}_{s}\right)
=
D_{\omega}(\mathbf{Z})
=
\left(
H_f(\mathbf{Z}),
H_s(\operatorname{Pool}_{\pi}(\mathbf{Z}))
\right),
\end{equation}
where $\mathbf{C}_{f}\in\mathbb{R}^{L\times D}$ and
$\mathbf{C}_{s}\in\mathbb{R}^{S\times D}$. Here, $\pi$ denotes the segment
assignment learned from music self-similarity. Unlike fixed temporal pooling,
the two representations are learned jointly with the downstream choreography
objectives. $\mathbf{C}_{s}$ captures slowly varying musical structure relevant
to formation and transition planning, whereas $\mathbf{C}_{f}$ retains local
rhythmic cues required for detailed motion synthesis and beat alignment.

\vspace{-2pt}
\textbf{Controllable macro-level intent.} A frozen omni-modal model
$\Phi_{\mathrm{omni}}$~\citep{Qwen3-Omni} receives the music audio and sparse
attributes such as music style, dance style, and group size. It produces a
description $d$ of macro-level properties including mood, energy progression,
movement intensity, unison, and formation intent:
\begin{equation}
d
=
\Phi_{\mathrm{omni}}(\mathbf{a},\mathbf{y}),
\end{equation}
where $\mathbf{a}$ denotes the music audio and $\mathbf{y}$ the sparse attributes.
A frozen Qwen3-Embedding-0.6B text encoder~\citep{zhang2025qwen3} converts $d$ into semantic tokens:
\begin{equation}
\mathbf{E}_{d}
=
\Phi_{\mathrm{text}}(d)
\in
\mathbb{R}^{Q\times D_e}.
\end{equation}
This yields two complementary planning signals: $\mathbf{C}_{s}$ describes the learned temporal structure of the music, while $\mathbf{E}_{d}$ supplies interpretable semantic intent. Both condition the Director, whereas $\mathbf{C}_{f}$ is reserved for frame-level motion realization.

\vspace{-6pt}
\subsection{Explicit Choreography Director}
\label{sec:director}

\vspace{-4pt}
The group-dance Director maps musical structure, semantic intent, and the current group state to an explicit choreography plan $\mathbf{P}$. Its computation is organized into four stages, i.e., \emph{Perceive}, \emph{Plan}, \emph{Schedule}, and \emph{Encode}, so that spatial organization, transition timing, and dancer participation remain explicit rather than being hidden inside the motion generator.

%The group-dance Director transforms segment-level planning context and the
%current group state into an explicit choreography plan. It follows four stages:
%\emph{Perceive}, \emph{Plan}, \emph{Schedule}, and \emph{Encode}.

\vspace{-4pt}
\textbf{State-Aware Perception.} At cycle $r\in\{1,\ldots,T\}$, the state encoder maps the group state $\mathcal{S}^{(r)}$ and an unknown-motion mask to dancer--segment tokens. Here, $\mathcal{S}^{(r)}$ summarizes available root geometry and motion, while the mask distinguishes valid dancers with unavailable motion from padded dancers. The feedback encoder maps $\mathbf{F}^{(r-1)}$ to diagnostic tokens, and the semantic adapter projects $\mathbf{E}_{d}$ into the Director space. The state tokens then query these contexts together with segment-level music $\mathbf{C}_{s}$ through cross-attention. At $r=1$, all valid dancers are marked motion-unknown and an empty-feedback token is used; for $r>1$, the generated motion from the previous cycle provides the observable group state.
 
To distinguish dancers before motion is available while preserving permutation
equivariance, we independently sample a centered actor seed for each valid
dancer and fix it throughout the sample:
\vspace{-2pt}
\begin{equation}
\mathbf{z}_i \sim \mathcal{N}(\mathbf{0},\mathbf{I}),
\qquad
\tilde{\mathbf{z}}_i
=
\mathbf{z}_i
-
\frac{1}{N}\sum_j\mathbf{z}_j.
\end{equation}
The seeds permute with the dancers, distinguishing actors without assigning roles to arbitrary dancer indices; the Director derives their context-dependent spatial roles from the planning context. A two-layer MLP state encoder combines the current group state and actor seed:
\begin{equation}
\mathbf{h}^{(r,0)}_{i,s}
=
E_{\mathrm{state}}
\left(
[\bar{\mathcal{S}}^{(r)}_{i,s};
\tilde{\mathbf{z}}_i]
\right),
\end{equation}
where $\bar{\mathcal{S}}^{(r)}_{i,s}$ replaces unavailable motion with a
learned unknown-motion embedding. Actor seeds distinguish exchangeable dancers
but encode no predefined roles, which the Director predicts through
$\boldsymbol{\rho}^{(r)}$. The tokens query music, semantics, and feedback through cross-attention with an active residual, exposing the Director to these contexts from the start of training:
\begin{equation}
\mathbf{H}^{(r)}
\operatorname{Fusion}
\left(
\mathbf{H}^{(r,0)},
[\mathbf{C}{s};\mathbf{E}{d};\mathbf{E}_{\mathrm{fb}}^{(r)}]
\right).
\end{equation}

\begin{figure*}[t]
    \centering
    \includegraphics[width=0.88\textwidth]{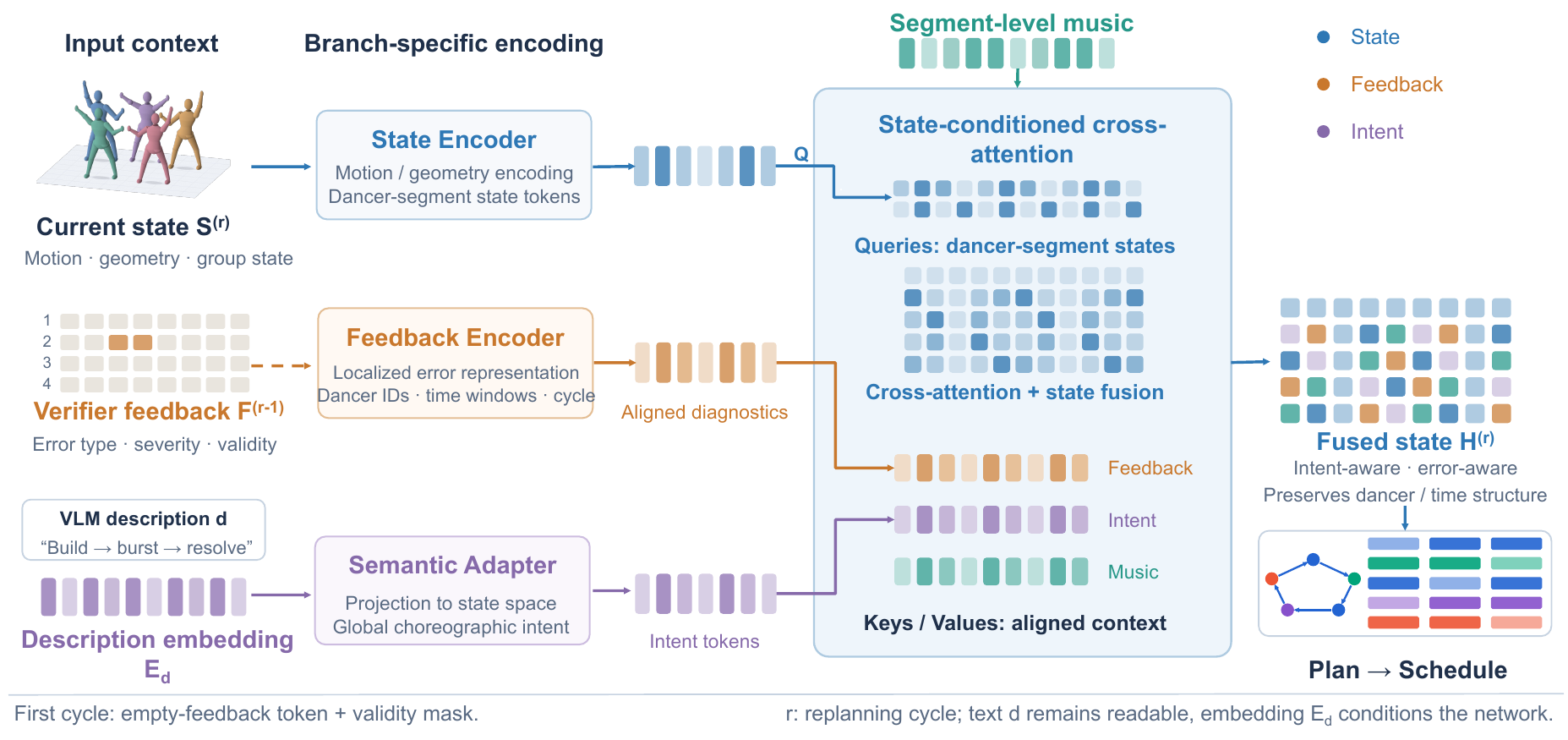}
    \caption{\textbf{State-aware perception in the Director.} Exchangeable actor seeds distinguish dancers without assigning roles to their indices, while later cycles use the generated geometry and motion state. State tokens attend to segment-level music, the high-level description, and optional feedback to produce dancer-segment representations for choreography planning.}
    \label{fig:perception}
\end{figure*}

\vspace{-2pt}
\textbf{Relative Formation Planning.} For each segment $s$, the Director constructs a multi-head directed relation graph. The edge from dancer $i$ to $j$ in head $h$ is:
\begin{equation}
A^{h}_{ij,s}
=
\operatorname{softmax}_{j}
\left(
f_h
\left[
W_q^h\mathbf{h}_{i,s},
W_k^h\mathbf{h}_{j,s},
\mathbf{e}_{ij,s},
\bar{\mathbf{h}}_s
\right]
\right),
\end{equation}
where $\mathbf{e}_{ij,s}$ contains available relative geometric features and $\bar{\mathbf{h}}_s$ is the masked group representation. Separate query and key projections allow $\mathbf{A}$ to represent asymmetric relationships. Graph message passing~\citep{gilmer2017neural} produces a formation proposal $\tilde{\mathbf{Y}}_{i,s}\in\mathbb{R}^{2}$ for each dancer. We remove the group centroid and separate formation shape from group scale:
\begin{align}
\mathbf{Q}_{i,s}
&=
\tilde{\mathbf{Y}}_{i,s}
-
\frac{1}{N}\sum_j\tilde{\mathbf{Y}}_{j,s},
\\
\mathbf{Y}_{i,s}
&=
\alpha_s
\frac{\mathbf{Q}_{i,s}}
{
\sqrt{
\frac{1}{N}
\sum_j
\|\mathbf{Q}_{j,s}\|_2^2
+\epsilon
}},
\end{align}
where $\alpha_s$ is the predicted metric RMS centroid radius. Pairwise pairwise formation geometry is:
\begin{equation}
D_{ij,s}
=
\left\|
\mathbf{Y}_{i,s}
-
\mathbf{Y}_{j,s}
\right\|_2.
\end{equation}
Because pairwise distances are derived from a common 2D point set rather than predicted independently, the resulting distance matrix is geometrically realizable by construction.

We derive a soft spatial-role score from the incoming mass of the directed relation graph:
\begin{equation}
\rho_{i,s}
=
\frac{1}{H}
\sum_h\sum_j A^h_{ji,s}
-
\frac{1}{N}
\sum_k
\frac{1}{H}
\sum_h\sum_j A^h_{jk,s}.
\end{equation}
We couple relation reasoning and formation planning instead of treating them as
independent branches. After a graph update, the Director drafts a point set and
derives its pairwise distances $\tilde{D}_{ij,s}$. These distances are projected
back into the edge representation before the next graph update:
\begin{equation}
\mathbf{e}_{ij,s}^{(\ell+1)}
=
\mathbf{e}_{ij,s}^{(\ell)}
+
W_D\tilde{D}_{ij,s}.
\end{equation}
The relation graph therefore reasons about the geometry it proposes, while the final formation depends on the dancer-specific relational context.

\vspace{-4pt}
\textbf{Coupled Transition and Mobility Scheduling.} A spatial target specifies the desired configuration but not \emph{when} it should change or \emph{which dancers} should carry the transition. We therefore predict a segment-level transition gate:
\begin{equation}
g_s
=
\sigma
\left(
f_g[
\bar{\mathbf{h}}_s,
\Delta_s^{\mathrm{plan}},
\Delta_s^{\mathrm{state}},
\eta_s
]
\right),
\end{equation}
where $\Delta_s^{\mathrm{plan}}$ measures the change between consecutive planned formations, $\Delta_s^{\mathrm{state}}$ measures the change currently observed in generated motion, and $\eta_s$ indicates how much motion state is known. The group representation $\bar{\mathbf{h}}_s$ is already conditioned on segment-level music and the high-level description. The gate converts the raw formation proposals into a temporally scheduled plan:
\begin{equation}
\mathbf{Y}^{\mathrm{sch}}_{i,s}
=
(1-g_s)\mathbf{Y}^{\mathrm{sch}}_{i,s-1}
+
g_s\mathbf{Y}_{i,s}.
\end{equation}
The scheduled distances $D^{\mathrm{sch}}_{ij,s}$ and scale
$\alpha^{\mathrm{sch}}_s$ are recomputed from
$\mathbf{Y}^{\mathrm{sch}}_{s}$.
Thus, a low gate preserves the previous relative formation while still allowing local body motion, whereas a high gate activates a formation transition. For an activated transition, the Director predicts the mobility allocation:
\begin{equation}
\mu_{i,s}
=
\operatorname{softmax}_{i}
\left(
f_{\mu}
[
\mathbf{h}_{i,s},
\left\|
\mathbf{Y}^{\mathrm{sch}}_{i,s}
-
\mathbf{Y}^{\mathrm{sch}}_{i,s-1}
\right\|_2
]
\right).
\end{equation}
The resulting $\boldsymbol{\mu}_s$ determines which dancers execute the transition and how movement is distributed among them. Consequently, formation geometry, transition timing, and dancer participation are jointly determined: the proposed geometry affects the gate, and the scheduled displacement in turn determines the mobility allocation.

\vspace{-4pt}
\textbf{Explicit Plan Encoding.}
The generator receives group-level choreography only through the structured
plan. Since formation orientation is arbitrary, each dancer instruction encodes
rotation-invariant geometry together with the scheduled plan variables:
\begin{equation}
\mathbf{u}_{i,s}
=
E_{\mathbf{P}}
\left(
\mathbf{D}^{\mathrm{sch}}_{i,:,s},
\left\|\mathbf{Y}^{\mathrm{sch}}_{i,s}\right\|_2,
g_s,
\mu_{i,s},
\rho_{i,s},
\alpha^{\mathrm{sch}}_s,
\delta_s
\right),
\end{equation}
where $\delta_s$ controls inter-dancer pose divergence. The segment-level
instructions are linearly interpolated into frame-aligned tokens
$\mathbf{U}\in\mathbb{R}^{N\times L\times D_u}$ to avoid boundary
discontinuities. The directed relation graph provides an additive bias for cross-dancer spatial attention:
\begin{equation}
B^{h}_{ij,t}
=
\kappa_{\pi(t)}
\log
\left(
A^{h}_{ij,\pi(t)}+\epsilon
\right),
\end{equation}
where $\kappa_s$ controls interaction strength. The bias affects only spatial
attention across dancers, leaving temporal attention and music conditioning
unchanged. At cycle $r$, the two plan conditions are denoted by
$\mathbf{U}^{(r)}$ and $\mathbf{B}^{(r)}$. Crucially, these are the only Director outputs exposed to the generator; no unrestricted hidden state bypasses the explicit plan interface. This restriction prevents implicit bypasses, making the plan directly interpretable, editable, and testable.

\vspace{-5pt}
\subsection{Plan-Conditioned Group Dance Generation}
\label{sec:generation}

\vspace{-4pt}
Given a fixed plan for the current cycle, the Dance Generator jointly synthesizes all dancers with a conditional diffusion model. At diffusion step $k$ of refinement cycle $r$, it receives noisy group
motion $\mathbf{X}^{(r)}_k$, frame-level rhythmic condition
$\mathbf{C}_{f}$, frame-aligned dancer instruction tokens
$\mathbf{U}^{(r)}$, and relation-derived cross-dancer spatial attention bias
$\mathbf{B}^{(r)}$:
\begin{equation}
\hat{\mathbf{X}}^{(r)}_0
=
G_{\phi}
\left(
\mathbf{X}^{(r)}_k,
k,
\mathbf{C}_{f},
\mathbf{U}^{(r)},
\mathbf{B}^{(r)}
\right).
\end{equation}
The model directly predicts the clean motion, and the completed diffusion trajectory yields candidate $\mathbf{X}^{(r)}$. The instruction tokens modulate every transformer block through token-wise adaptive layer normalization. Meanwhile, $\mathbf{B}^{(r)}$ is added to the logits of cross-dancer attention:
\begin{equation}
\operatorname{Attn}
(\mathbf{Q},\mathbf{K},\mathbf{V})
=
\operatorname{softmax}
\left(
\frac{\mathbf{Q}\mathbf{K}^{\top}}{\sqrt{D_h}}
+
\mathbf{B}^{(r)}
\right)
\mathbf{V}.
\end{equation}
The two conditioning paths serve complementary roles: $\mathbf{U}^{(r)}$ specifies dancer-wise plan variables, while $\mathbf{B}^{(r)}$ directly shapes cross-dancer information exchange during synthesis.

We sample diffusion levels per dancer and mix in per-token noise levels during training. This exposes the model to partially observed motion and allows known tokens, such as an overlap with a preceding generation window, to remain fixed during long-sequence generation.

\vspace{-5pt}
\subsection{Verification-Guided Refinement}
\label{sec:refinement}

\vspace{-4pt}
A single Director--Generator pass may fail to realize the intended choreography or may introduce physical artifacts. ChoreoStudio therefore performs up to $T$ closed-loop refinement cycles, using deterministic verification to convert generation failures into structured feedback for replanning. After trajectory $r$, the frozen Verifier evaluates $\mathbf{X}^{(r)}$ with $\mathbf{P}^{(r)}$ and $\mathbf{C}_{f}$ using deterministic plan-realization, collision, spacing, dispersion, foot-skating, freezing, and beat-alignment checks:
\begin{equation}
\left(
R^{(r)},b^{(r)},\mathbf{F}^{(r)}
\right)
=
V
\left(
\mathbf{X}^{(r)},
\mathbf{P}^{(r)},
\mathbf{C}_{f}
\right),
\end{equation}
where $R^{(r)}$, $b^{(r)}$, and $\mathbf{F}^{(r)}$ denote the aggregate reward,
pass flag, and localized feedback. The next-cycle inputs are:
\begin{equation}
\mathbf{E}_{\mathrm{fb}}^{(r+1)}
=
E_{\mathrm{fb}}\left(\mathbf{F}^{(r)}\right),
\qquad
\mathcal{S}^{(r+1)}
=
\operatorname{Summarize}\left(\mathbf{X}^{(r)}\right),
\end{equation}
which condition the Director when predicting $\mathbf{P}^{(r+1)}$. Refinement therefore changes the explicit plan rather than directly editing the generated motion. If a candidate passes, refinement may terminate early; otherwise, after at most $T$ cycles, we return the highest-reward candidate, prioritizing passed candidates when available. Verifier details are provided in Appendix~\ref{app:verifier}.

\vspace{-4pt}
\subsection{Training Objectives}
\label{sec:objectives}

\vspace{-2pt}
All supervision for the explicit plan is derived automatically from ground-truth motion, requiring no manual choreography annotation. We match the exchangeable actor seeds to target trajectories once per sequence and preserve this assignment across segments, preventing arbitrary dancer order from defining roles. To avoid a moving-target problem between planning and generation, training decouples plan prediction from plan execution. The Dance Generator is trained to realize a teacher plan extracted from ground-truth motion, whereas the Director learns to recover that plan from segment-level music, semantic intent, and the available group state.
\begin{equation}
\begin{aligned}
\mathcal{L}
={}&
\mathcal{L}_{\mathrm{motion}}
+\lambda_{\mathrm{dir}}\mathcal{L}_{\mathrm{dir}}
+\lambda_{\mathrm{real}}\mathcal{L}_{\mathrm{real}}
+\lambda_{\mathrm{space}}\mathcal{L}_{\mathrm{space}}
+\lambda_{\mathrm{jump}}\mathcal{L}_{\mathrm{jump}}.
\end{aligned}
\end{equation}
Here, $\mathcal{L}_{\mathrm{motion}}$ supervises motion reconstruction and
dynamics, $\mathcal{L}_{\mathrm{dir}}$ plan prediction, and
$\mathcal{L}_{\mathrm{real}}$ plan realization, while
$\mathcal{L}_{\mathrm{space}}$ and $\mathcal{L}_{\mathrm{jump}}$ discourage
formation collapse and implausible root displacement, respectively. Full
definitions are provided in Appendix~\ref{app:objectives}.

\section{Experiments}
\label{sec:experiments}

\vspace{-4pt}
\subsection{Experimental Setup}
\label{sec:exp_setup}

\vspace{-2pt}
We evaluate ChoreoStudio on AIOZ-GDance~\citep{le2023music}, using its official split of 1,309 training and 154 test sequences. A single model is trained on all group sizes from two to seven dancers. Full data processing and motion representations are provided in Appendix~\ref{app:implementation}.

\vspace{-2pt}
\textbf{Baselines and protocol.} We compare with GCD~\citep{le2023controllable}, CoDancers~\citep{yang2024codancers}, TCDiff~\citep{dai2025harmonious}, TCDiff++~\citep{dai2026tcdiffpp}, and ST-GDance~\citep{xu2025stgdance}. All methods use the same test identities, sequence length, motion conversion, feature extractors, and evaluation code; published scores from incompatible protocols are not mixed with our results.

\vspace{-2pt}
\textbf{Metrics.} We report motion fidelity and diversity, PFC (Physical Foot Contact)~\citep{tseng2023edge}, GMR (Group Motion Realism), GMC (Group Motion Correlation), TIF (Trajectory Intersection Frequency). IDDV (Inter-Dancer Distance Variance), Jitter, and FMR (Formation Meaningfulness Ratio) measure the low- and high-frequency components of formation dynamics. BAS (Beat Alignment Score)~\citep{siyao2022bailando} measures the temporal alignment between kinematic beats and musical beats. Complete definitions are in Appendix~\ref{app:metrics}.

\vspace{-2pt}
\textbf{Implementation.} Main results use the 150-frame setting, with a 720-frame evaluation in Appendix~\ref{app:long_sequence}. The same ChoreoStudio model is evaluated at both lengths and across all dancer counts without retraining, whereas each baseline is trained separately for each group size. During inference, the Director revises its plan using localized feedback until a valid candidate reaches the Verifier score threshold or the cycle budget is exhausted; the highest-scoring candidate is returned. Optimization, sampling, refinement budgets, and hardware settings are provided in Appendix~\ref{app:implementation}.

\vspace{-6pt}
\subsection{Comparison with Existing Methods}
\label{sec:comparison}

\vspace{-2pt}
\textbf{Quantitative comparison.} Table~\ref{tab:main_results} reports single- and multi-dancer performance under the shared 150-frame protocol. ChoreoStudio attains the best GMR and GMC at every group size, indicating that explicit planning improves group-level organization rather than only local motion quality, and the margin does not degrade as the group grows. Prior methods trade these properties against each other: CoDancers achieves low collision rates but the weakest group realism and correlation, consistent with avoiding intersections by keeping dancers apart, whereas TCDiff++ and ST-GDance improve realism without closing the gap in coordination. ChoreoStudio instead improves both simultaneously while achieving the best or second-best TIF across group sizes, showing that formation planning does not
compromise collision avoidance.

\begin{table*}[t]
\centering
\vspace{-8pt}
\caption{\textbf{Single and Group dance performance on AIOZ-GDance.} Best
results are \textbf{bold}; second-best results are \underline{underlined}.}
\label{tab:main_results}
\setlength{\tabcolsep}{4.5pt}
\renewcommand{\arraystretch}{1.12}
\resizebox{\textwidth}{!}{
\begin{tabular}{l|ccc|ccc|ccc|ccc}
\hline
\multirow{2}{*}{Method} &
\multicolumn{3}{c|}{\textbf{1 Dancer}} &
\multicolumn{3}{c|}{\textbf{2 Dancers}} &
\multicolumn{3}{c|}{\textbf{3 Dancers}} &
\multicolumn{3}{c}{\textbf{4 Dancers}} \\
\cline{2-13}
& FID$\downarrow$ & Div$\uparrow$ & PFC$\downarrow$
& GMR$\downarrow$ & GMC$\uparrow$ & TIF$\downarrow$
& GMR$\downarrow$ & GMC$\uparrow$ & TIF$\downarrow$
& GMR$\downarrow$ & GMC$\uparrow$ & TIF$\downarrow$ \\
\hline
GCD~\cite{le2023controllable}
& 39.24 & 9.64 & 2.53
& 34.39 & 80.32 & 0.17
& 30.22 & 80.22 & 0.19
& 36.28 & \underline{81.82} & 0.13 \\

CoDancers~\cite{yang2024codancers}
& 23.98 & 9.48 & 3.53
& 24.55 & 72.52 & \textbf{0.08}
& 26.34 & 74.22 & \textbf{0.08}
& 26.44 & 75.34 & \underline{0.10} \\

TCDiff++~\cite{dai2026tcdiffpp}
& \underline{22.42} & \textbf{17.43} & 1.32
& \underline{15.77} & \underline{80.87} & 0.11
& \underline{11.57} & 82.31 & \underline{0.10}
& \underline{13.63} & 81.56 & 0.13 \\

ST\mbox{-}GDance~\cite{xu2025stgdance}
& 28.87 & 12.82 & \underline{0.97}
& 19.42 & 80.52 & 0.12
& 14.76 & 81.24 & 0.11
& 14.02 & 81.42 & \underline{0.10} \\
\hline
ChoreoStudio
& \textbf{22.34} & \underline{14.35} & \textbf{0.52}
& \textbf{13.56} & \textbf{81.64} & \underline{0.10}
& \textbf{10.42} & \textbf{83.23} & \textbf{0.08}
& \textbf{11.94} & \textbf{82.41} & \textbf{0.09} \\
\hline
\end{tabular}
}
\vspace{-2pt}
\end{table*}

\begin{wraptable}{r}{0.48\linewidth}
    \centering
    \vspace{-6pt}
    \caption{\textbf{Formation dynamics and music alignment score relative to real data.}}
    \label{tab:formation_dynamics}
    \scriptsize
    \setlength{\tabcolsep}{3pt}
    \resizebox{\linewidth}{!}{
        \begin{tabular}{lcccc}
            \hline
             & IDDV & Jitter & FMR & BAS \\
            \hline
            GT & 0.302 & 0.045 & 0.832 & 0.363 \\
            \hline
            Best baseline & 0.181 & 0.051 & 0.717 & 0.224 \\
            ChoreoStudio & 0.267 & 0.048 & 0.848 & 0.351 \\
            \hline
        \end{tabular}
    }
    \vspace{-8pt}
\end{wraptable}
\textbf{Formation dynamics and music alignment.} As shown in Table~\ref{tab:formation_dynamics}, ChoreoStudio consistently narrows the gap to real data across all four metrics. Compared with TCDiff++~\citep{dai2026tcdiffpp}, the strongest overall baseline, it recovers substantially more slow-timescale formation variation and stronger beat alignment while reducing high-frequency residuals. Its FMR remains close to the real-data reference, indicating that the recovered variation reflects structured formation evolution rather than rapid fluctuation.

\begin{figure*}[t]
    \centering
    \includegraphics[width=\textwidth]{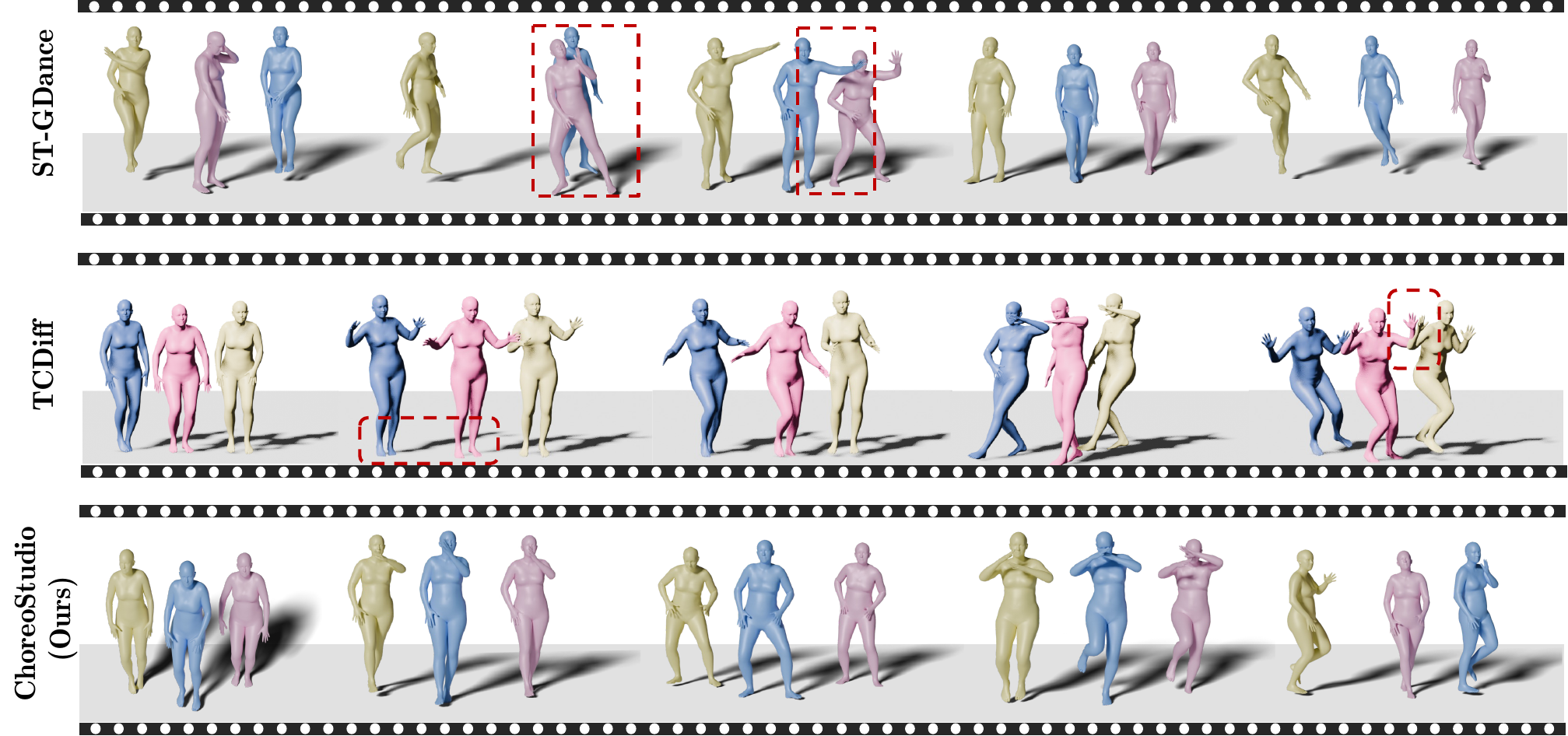}
    \caption{\textbf{Qualitative comparison on matched music segments.}
    We compare local motion and long-timescale formation evolution across
    methods using the same music and temporal extent.}
    \label{fig:qualitative_comparison}
\end{figure*}

\textbf{Qualitative comparison.}
Figure~\ref{fig:qualitative_comparison} compares ST-GDance~\citep{xu2025stgdance}, TCDiff~\citep{dai2025harmonious}, and ChoreoStudio from top to bottom under the same music. ST-GDance produces body overlap during position exchanges despite graph-based interaction modeling, while TCDiff exhibits foot-ground penetration and arm interpenetration. By planning formation targets, transition timing, and dancer-specific mobility before motion synthesis, ChoreoStudio coordinates spatial exchanges more reliably and produces stable full-body motion. Additional results are provided in Appendix~\ref{app:qualitative_results}.

\vspace{-6pt}
\subsection{Ablation Studies}
\label{sec:ablations}

\vspace{-4pt}
\begin{wraptable}{r}{0.48\linewidth}
    \centering
    \vspace{-8pt}
    \caption{\textbf{Ablation study for group dance generation (3 dancers).}}
    \label{tab:ablation}
    \scriptsize
    \setlength{\tabcolsep}{2.5pt}
    \resizebox{\linewidth}{!}{
        \begin{tabular}{lcccc}
            \hline
             & GMR$\downarrow$ & GMC$\uparrow$ & TIF$\downarrow$ & FMR$\uparrow$ \\
            \hline
            ChoreoStudio & \textbf{10.42} & \textbf{83.23} & \textbf{0.08} & \textbf{0.848} \\
            \hline
            w/o Semantic Description & 17.68 & 79.34 & 0.09 & 0.819 \\
            w/o Verifier & 16.10 & 78.68 & 0.10 & 0.803 \\
            w/o Director & 15.23 & 78.34 & 0.11 & 0.783 \\
            \hline
            w/o Role Assignment & 12.52 & 81.53 & 0.09 & 0.795 \\
            w/o Transition Timing & 14.32 & 79.32 & 0.10 & 0.803 \\
            w/o Relation Graph & 13.68 & 80.93 & 0.09 & 0.819 \\
            \hline
        \end{tabular}
    }
    \vspace{-8pt}
\end{wraptable}
All ablations degrade performance. Removing the Director causes the largest
drops in GMC and FMR and the highest TIF, supporting explicit choreography
planning. The semantic description improves group-motion realism, while the
Verifier strengthens coordination and formation dynamics. Among the structured
decisions, transition timing most affects realism and coordination, role
assignment contributes most to slow-timescale formation variation, and the
relation graph provides spatial organization. Together, these results support
jointly planning \emph{what} changes, \emph{when} it changes, and \emph{who}
executes the transition.

\vspace{-6pt}
\subsection{Plan Realization and Controlled Intervention}
\label{sec:intervention}

\vspace{-5pt}
\begin{wraptable}{r}{0.48\linewidth}
    \centering
    \vspace{-8pt}
    \caption{\textbf{Controlled plan interventions.} Error reduction measures improvements toward the edited target.}
    \label{tab:intervention}
    \scriptsize
    \setlength{\tabcolsep}{2.5pt}
    \renewcommand{\arraystretch}{1.28}
    \begin{tabular}{@{}lcccc@{}}
        \hline
        Intervention
        & \shortstack{Error reduction$\uparrow$}
        & Dir.$\uparrow$
        & $\Delta$TIF$\downarrow$
        & $\Delta$Jitter$\downarrow$ \\
        \hline
        Formation geometry & 0.40 & 78.54\% & 0.014 & 0.003 \\
        Transition timing    & 0.37 & 74.85\% & 0.018 & 0.004 \\
        Dancer mobility  & 0.29 & 68.68\% & 0.015 & 0.005 \\
        \hline
    \end{tabular}
    \vspace{-5pt}
\end{wraptable}

We perform paired interventions after plan prediction while fixing the music, actor seeds, diffusion noise, and non-intervened variables. Each edit is re-encoded before generation to isolate the plan component. As shown in Table~\ref{tab:intervention}, all three edits reduce realization error and change motion predominantly in the requested direction. Formation produces the strongest response, followed by timing, while mobility remains effective despite the constraint of preserving the target formation. The small changes in TIF and Jitter indicate that control is not achieved through motion degradation. Together, these results verify control over what changes, when it changes, and who executes the transition. Direct realization metrics and intervention controls are provided in Table~\ref{tab:plan_realization} and Appendix~\ref{app:plan_interventions}.

\vspace{-4pt}
\section{Conclusion}
\label{sec:conclusion}

\vspace{-4pt}

% We introduced ChoreoStudio to address a scale-specific gap in group dance generation: locally plausible motion can coexist with impoverished slow formation evolution. Its Director couples realizable relative geometry with transition timing and dancer mobility, then exposes the plan to full-body generation through dancer instructions and spatial attention biases. Experiments evaluate motion quality, formation dynamics, plan realization, and controlled interventions; absolute stage orientation and richer semantic roles remain future work.

We introduced ChoreoStudio to reveal and address a temporal-scale failure in
group dance generation: locally plausible motion can coexist with impoverished
slow formation evolution. Its Director jointly plans realizable relative
formations, transition timing, and dancer mobility, while an explicit
plan-to-motion interface makes these decisions executable, interpretable, and
editable. Experiments demonstrate state-of-the-art motion quality and group
organization, more realistic formation dynamics, accurate plan realization,
and causal responses to controlled interventions. Future work will incorporate
absolute stage orientation and richer semantic roles.

\subsection*{AI Use Statement}
In this work, we used generative AI tools for language polishing, grammar checking, and assistance with understanding and summarizing literature selected by the authors. These tools were not used for experimental implementation, data collection, or result analysis. All AI-assisted content was verified by the authors, who take responsibility for the final text, claims, and artifacts.

\subsection*{Ethics Statement}
All datasets used in this work are publicly available with established licenses. The user study involves anonymous participants providing perceptual rankings of generated dance sequences, with no collection of personally identifiable information.

\subsection*{Reproducibility Statement}

The architecture and training procedure are described in Section~\ref{sec:method}, and the evaluation protocol is presented in Section~\ref{sec:experiments}. Data processing, motion representation, semantic conditioning, state initialization, optimization, and sampling settings are detailed in Appendix~\ref{app:implementation}, with complete objective definitions in Appendix~\ref{app:objectives}. Evaluation metrics and statistical aggregation are specified in Appendix~\ref{app:metrics}. Controlled interventions and long-sequence generation are documented in Appendices~\ref{app:plan_interventions} and~\ref{app:long_sequence}, respectively. 

\bibliographystyle{plainnat}
\bibliography{iclr2027_conference}

\appendix

\section{Analysis and Metric Details}
\label{app:metrics}

This section provides the definitions and implementation details of all
evaluation metrics. Unless otherwise specified, generated and ground-truth
motions are converted to the same SMPL-based representation and evaluated with
the same feature extractors. Standard motion and group metrics are computed over
complete sequences. The frequency-based formation metrics are computed using
length-matched temporal windows and then aggregated at the sequence level.

\subsection{Motion Quality and Diversity}

Let $\phi_q(\mathbf{X})$ denote a motion feature extracted from sequence
$\mathbf{X}$, where $q\in\{k,g\}$ represents kinetic or geometric features. We
standardize real and generated features using the mean and standard deviation
of the real features. Let $(\boldsymbol{\mu}_r^q,\boldsymbol{\Sigma}_r^q)$ and
$(\boldsymbol{\mu}_g^q,\boldsymbol{\Sigma}_g^q)$ denote the Gaussian statistics
of the real and generated features, respectively. The corresponding Fr\'echet
distance is
\begin{equation}
\mathrm{FID}_q
=
\left\|
\boldsymbol{\mu}_r^q-\boldsymbol{\mu}_g^q
\right\|_2^2
+
\operatorname{Tr}
\left(
\boldsymbol{\Sigma}_r^q+
\boldsymbol{\Sigma}_g^q-
2\left(
\boldsymbol{\Sigma}_r^q\boldsymbol{\Sigma}_g^q
\right)^{1/2}
\right).
\end{equation}
Lower FID indicates that the generated feature distribution is closer to the
real distribution.

For $M$ generated sequences with standardized features
$\{\mathbf{z}_n^q\}_{n=1}^{M}$, diversity is their average pairwise distance:
\begin{equation}
\mathrm{Dist}_q
=
\frac{2}{M(M-1)}
\sum_{n<m}
\left\|
\mathbf{z}_n^q-\mathbf{z}_m^q
\right\|_2.
\end{equation}
We report the kinetic and geometric variants, denoted by FID$_k$/Dist$_k$ and
FID$_g$/Dist$_g$, respectively.

Physical Foot Contact (PFC) measures foot sliding during body acceleration.
Following the standard implementation, we compute the horizontal velocities of
the left and right feet, denoted by $v_L(t)$ and $v_R(t)$, and the normalized
root acceleration $\tilde{a}(t)$. The sequence-level score is
\begin{equation}
\mathrm{PFC}
=
\frac{10^4}{L-2}
\sum_{t=1}^{L-2}
\tilde{a}(t)\,v_L(t)\,v_R(t).
\end{equation}
For each foot, the smaller velocity of its toe and ankle joints is used. Lower
PFC indicates fewer foot-sliding artifacts.

\subsection{Group Motion Metrics}

Group Motion Realism (GMR) evaluates the discrepancy between real and generated
group-level kinetic-feature distributions. It uses the same Fr\'echet
formulation as above, replacing the individual-motion feature $\phi_q$ with a
group kinetic feature extractor $\phi_{\mathrm{grp}}$.

Group Motion Correlation (GMC) measures motion similarity between dancers. For
each dancer pair $(i,j)$, Dynamic Time Warping produces an alignment path
$\mathcal{A}_{ij}$ between their joint trajectories. Given frame-level kinetic
features $\mathbf{k}_i(t)$ and $\mathbf{k}_j(t')$, GMC is
\begin{equation}
\mathrm{GMC}
=
\frac{2}{N(N-1)}
\sum_{i<j}
\frac{1}{|\mathcal{A}_{ij}|}
\sum_{(t,t')\in\mathcal{A}_{ij}}
\mathbf{k}_i(t)^{\top}\mathbf{k}_j(t').
\end{equation}
Because excessively synchronized or duplicated motions may also produce a large
correlation, GMC is interpreted relative to its ground-truth value rather than
assuming that an arbitrarily larger value is always preferable.

Trajectory Intersection Frequency (TIF) measures inter-dancer body penetration.
We reconstruct the 24 SMPL joints and use the 23 bones defined by the SMPL
kinematic tree. Let $\mathcal{B}_i(t)$ denote the bone segments of dancer $i$ at
frame $t$. A dancer pair is considered intersecting if the minimum distance
between any two of their bone segments is below a thickness threshold $\delta$:
\begin{equation}
c_{ij}(t)
=
\mathbb{I}
\left[
\min_{\substack{b\in\mathcal{B}_i(t)\\
b'\in\mathcal{B}_j(t)}}
d_{\mathrm{seg}}(b,b')<\delta
\right].
\end{equation}
We use $\delta=0.10$\,m in the main experiments and compute
\begin{equation}
\mathrm{TIF}
=
\frac{1}{L}
\sum_{t=1}^{L}
\frac{2}{N(N-1)}
\sum_{i<j}c_{ij}(t).
\end{equation}
Thus, TIF is the average proportion of dancer pairs that intersect at each
frame. We apply this implementation uniformly to ground truth and all generated
results.

\subsection{Frequency-Based Formation Dynamics}
\label{app:frequency_metrics}

Static collision and correlation scores do not reveal whether formations
evolve meaningfully over time. We therefore decompose the temporal variation
of inter-dancer distances into gradual formation change and rapid residual
fluctuation.

Consider a group sequence with $N$ dancers and $L$ frames sampled at
$f_s=30$\,Hz. Let $\mathbf{p}_i(t)=(x_i(t),z_i(t))$ denote the ground-plane
root position of dancer $i$. For each unordered pair $(i,j)$, we construct
\begin{equation}
d_{ij}(t)=\left\|\mathbf{p}_i(t)-\mathbf{p}_j(t)\right\|_2.
\end{equation}
We apply a fourth-order zero-phase Butterworth low-pass filter with cutoff
$f_c=0.5$\,Hz:
\begin{equation}
d_{ij}^{\mathrm{low}}
=\operatorname{LP}_{f_c}[d_{ij}],
\qquad
d_{ij}^{\mathrm{high}}
=d_{ij}-d_{ij}^{\mathrm{low}}.
\end{equation}
The cutoff retains formation changes with characteristic periods of roughly
two seconds or longer while assigning rapid distance fluctuations to the
residual. The same filter is applied to real and generated sequences.

We divide each sequence into non-overlapping windows of $W=150$ frames. For
window $w$, define
\begin{align}
S_w
&=
\frac{2}{N(N-1)}
\sum_{i<j}
\operatorname{Std}_{t\in w}
\left[d_{ij}^{\mathrm{low}}(t)\right],\\
J_w
&=
\frac{2}{N(N-1)}
\sum_{i<j}
\sqrt{
\frac{1}{|w|}
\sum_{t\in w}
\left(d_{ij}^{\mathrm{high}}(t)\right)^2
}.
\end{align}
The final metrics are
\begin{equation}
\mathrm{IDDV}
=\frac{1}{|\mathcal W|}\sum_{w\in\mathcal W}S_w,
\qquad
\mathrm{Jitter}
=\frac{1}{|\mathcal W|}\sum_{w\in\mathcal W}J_w,
\end{equation}
and
\begin{equation}
\mathrm{FMR}
=
\frac{1}{|\mathcal W|}
\sum_{w\in\mathcal W}
\frac{S_w}{S_w+J_w+\epsilon}.
\end{equation}
IDDV measures slow formation evolution, Jitter measures rapid inter-dancer
distance fluctuation rather than joint-level pose jitter, and FMR measures the
fraction of variation attributable to gradual formation change. All three are
interpreted relative to real dance rather than optimized independently.

\subsection{Empirical Formation-Dynamics Comparison}
\label{app:formation_analysis}

\begin{figure*}[t]
    \centering
    \IfFileExists{figs/formation_analysis.pdf}{
        \includegraphics[width=\textwidth]{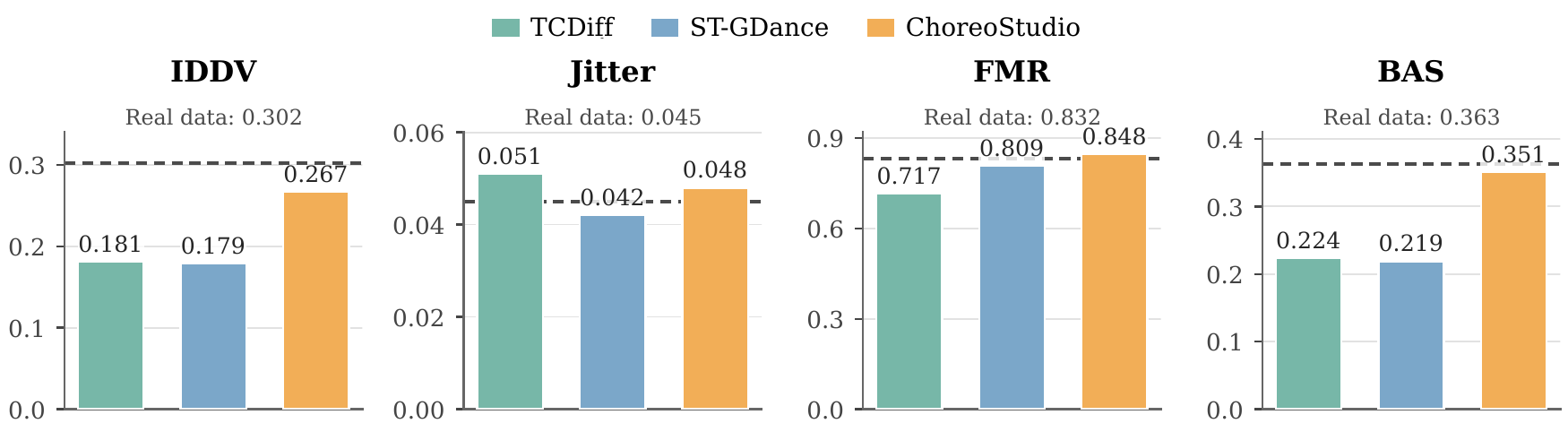}
    }{
        \fbox{
            \parbox[c][4.2cm][c]{0.96\textwidth}{
                \centering
                Placeholder for ground-plane trajectories, frequency-decomposed
                inter-dancer distances, and quantitative formation-dynamics
                comparisons.
            }
        }
    }
    \caption{\textbf{Formation dynamics and music alignment.} Bars report IDDV, Jitter, FMR, and BAS for TCDiff, ST-GDance, and ChoreoStudio. Dashed lines indicate the corresponding real-data references. All methods are evaluated using the same sequences, temporal windows, and metric implementations.}
    \label{fig:formation_analysis}
\end{figure*}

We apply the formation-dynamics metrics to ground-truth sequences and
representative group-dance generators under the same evaluation protocol.
Figure~\ref{fig:formation_analysis} reports the resulting dataset-level
statistics together with music alignment.

Both baselines substantially underrepresent slow formation evolution. TCDiff
and ST-GDance obtain IDDV scores of 0.181 and 0.179, respectively, compared
with 0.302 for real dance. TCDiff additionally exhibits stronger
high-frequency variation, with higher Jitter and substantially lower FMR than
the real-data references. ST-GDance produces more realistic Jitter and FMR but
retains a large IDDV gap. ChoreoStudio narrows this gap, achieving an IDDV of
0.267, Jitter of 0.048, and FMR of 0.848. Its BAS of 0.351 is also close to the
real-data value of 0.363, indicating that improved formation evolution is
obtained while preserving music alignment.

These metrics are interpreted relative to real data rather than optimized
independently. Excessive IDDV may indicate formation drift, while a high FMR
does not guarantee sufficient absolute formation change. Moreover, pairwise
distances omit formation orientation, dancer identity, and some interaction
semantics. The metrics therefore diagnose the temporal organization of
inter-dancer geometry rather than provide a complete measure of choreography
quality.

\subsection{Plan-Realization Metrics}

Let $\tilde{\mathbf{Y}}_i(t)$ denote the frame-level formation target obtained by interpolating the Director's segment-level plan, and define the generated root position relative to the group centroid as
\begin{equation}
\bar{\mathbf{p}}_i(t)
=
\mathbf{p}_i(t)
-
\frac{1}{N}\sum_j\mathbf{p}_j(t).
\end{equation}
Because the planned point set has no absolute stage orientation, we align it to the generated formation by a proper 2D rotation before measuring point-wise realization:
\begin{equation}
E_{\mathrm{form}}
=
\min_{\mathbf{R}\in\mathrm{SO}(2)}
\frac{1}{NL}
\sum_{i,t}
\left\|
\bar{\mathbf{p}}_i(t)-\mathbf{R}\tilde{\mathbf{Y}}_i(t)
\right\|_2.
\end{equation}
The actor-to-target correspondence is fixed by the sequence-level matching used during plan supervision. To evaluate relative geometry independently of translation and rotation, we also report the pairwise realization error
\begin{equation}
E_{\mathrm{pair}}
=
\frac{2}{N(N-1)L}
\sum_{i<j,t}
\left|
\left\|\mathbf{p}_i(t)-\mathbf{p}_j(t)\right\|_2
-
\left\|\tilde{\mathbf{Y}}_i(t)-\tilde{\mathbf{Y}}_j(t)\right\|_2
\right|.
\end{equation}

For the Director, transition decisions are evaluated using segment-level precision, recall, and F1 after selecting a gate threshold on the validation set. For each matched predicted and target transition, we also report the absolute temporal offset in seconds. These metrics distinguish whether the Director detects a transition from whether it places that transition at the correct time.
Mobility allocation is evaluated as the average $\ell_1$ distance between the
predicted and pseudo-label mobility distributions:
\begin{equation}
E_{\mathrm{mob}}
=
\frac{1}{S}
\sum_s
\left\|
\boldsymbol{\mu}_s-\boldsymbol{\mu}_s^*
\right\|_1.
\end{equation}
Lower values indicate better agreement for formation, pairwise geometry, mobility, and temporal-offset errors; larger transition F1 indicates better transition detection.

\begin{table}[H]
    \centering
    \caption{\textbf{Plan realization on the 150-frame three-dancer setting.}
    Formation and pairwise errors are measured in meters; lower is better
    except for transition F1.}
    \label{tab:plan_realization}
    \footnotesize
    \setlength{\tabcolsep}{4pt}
    \renewcommand{\arraystretch}{1.15}
    \begin{tabular}{lccccc}
        \hline
        Method
        & $E_{\mathrm{form}}\downarrow$
        & $E_{\mathrm{pair}}\downarrow$
        & Trans. F1$\uparrow$
        & Offset (s)$\downarrow$
        & $E_{\mathrm{mob}}\downarrow$ \\
        \hline
        ChoreoStudio & 0.13 & 0.09 & 0.80 & 0.32 & 0.16 \\
        \hline
    \end{tabular}
\end{table}

As shown in Table~\ref{tab:plan_realization}, ChoreoStudio closely realizes
the predicted choreography plan, with low formation and pairwise-geometry
errors. The transition F1 of 0.80 and average offset of 0.32 seconds indicate
that planned formation changes are generally detected and executed near their
scheduled times. The low mobility error further shows that dancer-specific
movement responsibilities are preserved during motion generation. Together,
these results confirm that the explicit plan is effectively translated into
formation geometry, transition timing, and dancer participation.

\subsection{Music-Dance Alignment}

Music-dance alignment is evaluated in the appendix using Beat Alignment Score
(BAS). Motion beats are identified as local minima of the total joint-velocity
envelope. Let $\mathcal{B}_x$ and $\mathcal{B}_m$ denote the detected motion
beats and input music beats, respectively. BAS is
\begin{equation}
\mathrm{BAS}
=
\frac{1}{|\mathcal{B}_x|}
\sum_{t\in\mathcal{B}_x}
\exp
\left(
-\frac{
\min_{b\in\mathcal{B}_m}|t-b|^2
}{2\sigma^2}
\right),
\end{equation}
where $\sigma=2.12$ frames following the standard evaluation protocol. The
score is first averaged across dancers and then across test sequences.

\subsection{Aggregation and Statistical Analysis}

For IDDV, Jitter, and FMR, metrics are first averaged across dancer
pairs within each temporal window, then across windows belonging to the same
sequence, and finally across test sequences. This prevents longer sequences
from receiving disproportionate weight. Comparisons with ground truth use
matched sequence identities whenever available.

\section{Implementation and Training Details}
\label{app:implementation}

\subsection{Data Processing and Motion Representation}

We use the official AIOZ-GDance~\citep{le2023music} training and test splits containing 1,309 and 154 sequences, respectively. Motion is sampled at $30\,\mathrm{fps}$. Each dancer is represented by a 151-dimensional vector containing three-dimensional root translation, 6D rotations for 24 SMPL joints, and four binary foot-contact states. The model is trained on all available group sizes from two to seven dancers. Within each batch, groups are padded to the largest dancer count and a binary validity mask is applied to dancer-wise attention, plan prediction, losses, and evaluation.

Training examples are extracted as 720-frame windows with a stride of 112 frames. The music disentangler identifies variable-length structural segments from the music representation. The Director predicts one choreography state for each learned segment, which is subsequently interpolated to the motion frame rate. For sequences longer than one generation window, inference uses 720-frame windows with a 30-frame overlap. Motion in the overlap is retained as known context during the next sampling window, and duplicated frames are removed after stitching.

\subsection{Music and Semantic Conditioning}

The acoustic input is a 438-dimensional feature sequence extracted offline at twice the motion frame rate. The music encoder produces frame-aligned rhythmic features for the Dance Generator and aggregates structural context over the learned music segments for the Director. Segment-level structural codes are trained from music self-similarity as described in Section~\ref{sec:conditioning}. Their contribution is evaluated by the \emph{w/o music structure} ablation in
Appendix~\ref{app:planning_diagnostics}.

The frozen omni-modal model receives the music audio and sparse attributes consisting of music style, dance style, and dancer count. Its description is generated once per sequence, encoded into semantic tokens, and cached before training. Gradients are not propagated through either the omni-modal model or its text encoder. The description supplies macro-level semantic intent, while transition timing and detailed motion remain grounded in the learned music features.

\subsection{Plan Targets and State Initialization}

Pseudo-label plans are extracted from ground-truth root trajectories without
manual choreography annotation. Root positions are centered by the group
centroid within each segment to define relative formation targets.
Scale-normalized changes in pairwise geometry determine transition labels,
while normalized dancer displacements define mobility targets. Complete
definitions are provided in Appendix~\ref{app:objectives}.

During training, the Director alternates between states derived from
ground-truth motion and initial states in which generated motion is unavailable.
In the latter case, independently sampled actor seeds distinguish otherwise
identical dancer tokens. The seeds are centered over valid dancers, permuted
with the dancer dimension, and fixed across all segments, generation windows,
and refinement cycles of the same sample. Padded dancers are excluded from seed
centering and all subsequent plan operations. Because the seeds assign no
semantics to dancer indices, we perform one sequence-level bipartite matching
between predicted actor plans and target dancer trajectories and preserve the
resulting correspondence over time.

\subsection{Verifier Implementation}
\label{app:verifier}

The frozen Verifier is applied once after each complete diffusion trajectory,
rather than at individual denoising steps. It receives the generated motion
$\mathbf{X}^{(r)}$, plan $\mathbf{P}^{(r)}$, and frame-level rhythmic condition
$\mathbf{C}_{f}$. World-coordinate joints
$\mathbf{J}\in\mathbb{R}^{N\times L\times J\times 3}$ are recovered through
forward kinematics, with the root given by $\mathbf{J}_{:,:,0,:}$.

\paragraph{Plan, physical, and coordination checks.}
The Verifier first measures plan realization using the discrepancy between the
scale-normalized pairwise geometry of generated roots and the scheduled plan:
\begin{equation}
c_{\mathrm{plan}}
=
\frac{1}{L}\sum_{t=1}^{L}
\left[
\frac{1}{N(N-1)}\sum_{i\neq j}
\left|
\widehat{D}^{X}_{ij,t}-\widehat{D}^{P}_{ij,t}
\right|
+\lambda_{\alpha}
\left|\alpha_t^X-\alpha_t^P\right|
\right].
\end{equation}
Here, $\alpha_t^X$ and $\alpha_t^P$ are the corresponding RMS centroid radii.
It additionally applies six deterministic checks. Capsule collision models each
skeleton bone as a capsule of radius $0.09$\,m and measures inter-dancer
penetration. Root spacing penalizes pairwise root distances below
$0.6$\,m. Group dispersion penalizes distances above $2.0$\,m, preventing
collision avoidance through excessive separation. This upper bound is
calibrated from real data, whose mean pairwise distance is $1.55$\,m and
90th percentile is approximately $2.0$\,m.

Foot skating is measured from horizontal foot velocity during contacts, where
contact requires both foot height and speed below $0.05$. The floor height is
estimated from the fifth percentile of foot heights in each sequence rather
than fixed globally. Anti-freeze detects dancers whose root and root-relative
joint velocities remain below $0.02$\,m/s within a one-second window. Beat
alignment compares musical beats with motion accents detected from joint
acceleration, using a Gaussian tolerance of three frames.

\paragraph{Reward and validity.}
The deterministic scores are aggregated as
\begin{equation}
R^{(r)}
=
c_{\mathrm{beat}}
-c_{\mathrm{plan}}
-2c_{\mathrm{coll}}
-c_{\mathrm{foot}}
-2c_{\mathrm{freeze}}
-c_{\mathrm{space}}
-c_{\mathrm{disp}}.
\end{equation}
The hard validity flag is computed independently:
\begin{equation}
b^{(r)}
=
\mathbb{I}
\left[
c_{\mathrm{coll}}\leq0.02,\;
c_{\mathrm{freeze}}\leq0.25,\;
c_{\mathrm{space}}\leq0.05,\;
c_{\mathrm{disp}}\leq0.15
\right].
\end{equation}
Thus, $R^{(r)}$ ranks candidates, whereas $b^{(r)}$ rejects motions with severe
physical or spatial violations.

\paragraph{Localized feedback.}
Each violation records its type, normalized start and end times, severity, and
affected dancers. Events are ranked by severity, and at most eight are retained
per sample. They are encoded as event features, dancer masks, and verifier
sub-scores to form $\mathbf{F}^{(r)}$. The dispersion cost affects reward and
candidate selection but is not appended to the feedback score vector, preserving
compatibility with the trained feedback encoder; plan mismatches enter through
localized event features.

\paragraph{Feedback-branch training.}
During training, the frozen Verifier evaluates completed motion samples and
converts localized violations into feedback tokens. These tokens and the
summarized motion state condition a subsequent Director prediction, which is
supervised by the same plan targets through $\mathcal L_{\mathrm{dir}}$.
Gradients update the feedback encoder and Director but do not pass through the
Verifier or diffusion sampling. Empty-feedback examples are retained to train
the first cycle.

\paragraph{Candidate selection.}
The refinement memory stores the plan, motion, reward, validity, and feedback
from every completed cycle. Refinement terminates once a valid candidate
reaches the validation-set reward threshold $\tau_R$ or the maximum of $T$
cycles. The system
returns the highest-reward passed candidate among the completed cycles; if none
passes, it returns the highest-reward candidate with its diagnostics. Main
results use this multi-cycle selection protocol.

\subsection{Detailed Objective Definitions}
\label{app:objectives}

For each segment $s$, ground-truth root positions are averaged temporally and centered by the group centroid to obtain $\mathbf{Y}^{*}_{i,s}$. The target pairwise-distance matrix is $D^{*}_{ij,s}=\|\mathbf{Y}^{*}_{i,s}-\mathbf{Y}^{*}_{j,s}\|_2$. We define the target scale using the same RMS centroid radius as the Director,
\begin{equation}
\alpha_s^{*}
=
\sqrt{\frac{1}{N}\sum_i\|\mathbf{Y}^{*}_{i,s}\|_2^2+\epsilon},
\end{equation}
and $\hat{D}^{*}_{ij,s}=D^{*}_{ij,s}/(\alpha_s^{*}+\epsilon)$ defines scale-free formation shape.

Actor seeds do not prescribe a correspondence with dataset dancer indices. We therefore compute one matching over the complete segment sequence using rotation- and permutation-invariant geometric signatures. Let $r_{i,s}=\|\mathbf{Y}_{i,s}\|_2$ and let $\operatorname{sort}(\mathbf{D}_{i,:,s})$ denote the sorted distances from dancer $i$ to the other dancers. The matching cost is
\begin{equation}
C_{ij}=\frac{1}{S}\sum_s\left|r_{i,s}-r^{*}_{j,s}\right|+\lambda_D\frac{1}{S}\sum_s\left\|\operatorname{sort}(\mathbf{D}_{i,:,s})-\operatorname{sort}(\mathbf{D}^{*}_{j,:,s})\right\|_1+\lambda_{\mathrm{mob}}\frac{1}{S}\sum_s\left|\mu_{i,s}-\mu^{*}_{j,s}\right|.
\end{equation}
Hungarian matching produces a permutation $\Pi^{*}$ that is held fixed for all segments. All dancer-indexed targets below are reordered by $\Pi^{*}$ before computing losses.

The transition target detects a sufficiently large change in scale-normalized pairwise geometry:
\begin{equation}
g_s^{*}=\mathbb{I}\left[\Delta\left(\mathbf{D}^{*}_{s},\mathbf{D}^{*}_{s-1}\right)>\tau_g\right],\qquad s\geq2.
\end{equation}
The target mobility distribution measures the relative displacement required from each dancer:
\begin{equation}
\mu^{*}_{i,s}=\frac{\left\|\mathbf{Y}^{*}_{i,s}-\mathbf{Y}^{*}_{i,s-1}\right\|_2}{\sum_j\left\|\mathbf{Y}^{*}_{j,s}-\mathbf{Y}^{*}_{j,s-1}\right\|_2+\epsilon},\qquad s\geq2.
\end{equation}
This target captures formation-level travel rather than local in-place body motion.

The Director objective is
\begin{equation}
\mathcal{L}_{\mathrm{dir}}=\lambda_{\mathrm{shape}}\mathcal{L}_{\mathrm{shape}}+\lambda_{\mathrm{scale}}\mathcal{L}_{\mathrm{scale}}+\lambda_{\mathrm{gate}}\mathcal{L}_{\mathrm{gate}}+\lambda_{\mathrm{mob}}\mathcal{L}_{\mathrm{mob}}+\lambda_{\mathrm{div}}\mathcal{L}_{\mathrm{div}}+\lambda_{\mathrm{rate}}\mathcal{L}_{\mathrm{rate}}.
\end{equation}
Its components are
\begin{align}
\mathcal{L}_{\mathrm{shape}}&=\frac{1}{N(N-1)S}\sum_{i\neq j,s}\left|\hat{D}_{ij,s}-\hat{D}^{*}_{ij,s}\right|,\\
\mathcal{L}_{\mathrm{scale}}&=\frac{1}{S}\sum_s\left|\alpha_s-\alpha_s^{*}\right|,\\
\mathcal{L}_{\mathrm{gate}}&=-\frac{1}{S-1}\sum_{s=2}^{S}\left[g_s^{*}\log g_s+(1-g_s^{*})\log(1-g_s)\right],\\
\mathcal{L}_{\mathrm{mob}}&=\frac{\sum_{s=2}^{S}g_s^{*}\left\|\boldsymbol{\mu}_s-\boldsymbol{\mu}_s^{*}\right\|_1}{\sum_{s=2}^{S}g_s^{*}+\epsilon},\\
\mathcal{L}_{\mathrm{div}}&=\frac{1}{S}\sum_s\left|\delta_s-\delta_s^{*}\right|,\\
\mathcal{L}_{\mathrm{rate}}&=\left|\frac{1}{S-1}\sum_{s=2}^{S} g_s-\frac{1}{S-1}\sum_{s=2}^{S} g_s^{*}\right|.
\end{align}
We do not directly supervise the directed relation graph $\mathbf{A}$ because no corresponding annotation is available. Transition and mobility targets are defined from $s=2$ onward; mobility supervision is applied only when $g_s^*=1$.

Let $\hat{\mathbf{X}}_0$ be the predicted clean motion and $\mathbf{X}_0$ the ground truth. The motion objective is
\begin{equation}
\mathcal{L}_{\mathrm{motion}}=\lambda_{\mathrm{rec}}\mathcal{L}_{\mathrm{rec}}+\lambda_{\mathrm{vel}}\mathcal{L}_{\mathrm{vel}}+\lambda_{\mathrm{pos}}\mathcal{L}_{\mathrm{pos}}+\lambda_{\mathrm{contact}}\mathcal{L}_{\mathrm{contact}},
\end{equation}
where
\begin{align}
\mathcal{L}_{\mathrm{rec}}&=\frac{1}{NL}\sum_{i,t}\left\|\hat{\mathbf{x}}_{i,t}-\mathbf{x}_{i,t}\right\|_2^2,\\
\mathcal{L}_{\mathrm{vel}}&=\frac{1}{N(L-1)}\sum_{i,t}\left\|\Delta\hat{\mathbf{x}}_{i,t}-\Delta\mathbf{x}_{i,t}\right\|_2^2,\\
\mathcal{L}_{\mathrm{pos}}&=\frac{1}{NLJ}\sum_{i,t,j}\left\|\operatorname{FK}(\hat{\mathbf{x}}_{i,t})_j-\operatorname{FK}(\mathbf{x}_{i,t})_j\right\|_2^2,\\
\mathcal{L}_{\mathrm{contact}}&=\frac{1}{NL|\mathcal{F}|}\sum_{i,t,f}c^{*}_{i,t,f}\left\|\mathbf{v}^{xz}_{i,t,f}\right\|_2.
\end{align}
Here, $J$ is the number of joints, $\mathcal{F}$ is the set of foot joints, $c^{*}_{i,t,f}$ is the ground-truth contact indicator, and $\mathbf{v}^{xz}_{i,t,f}$ is the generated horizontal foot velocity. The contact term is evaluated on ground-truth support frames.

To supervise execution of the conditioning plan, we pool the generated root trajectories to the same segment grid, remove the group centroid, and compute their scale-free pairwise shape $\hat{D}^{\mathrm{gen}}_{ij,s}$ and scale $\alpha^{\mathrm{gen}}_s$. The realization loss is
\begin{equation}
\mathcal{L}_{\mathrm{real}}=\frac{1}{N(N-1)S}\sum_{i\neq j,s}\left|\hat{D}^{\mathrm{gen}}_{ij,s}-\hat{D}^{\mathrm{plan}}_{ij,s}\right|+\lambda_{\alpha}\frac{1}{S}\sum_s\left|\alpha^{\mathrm{gen}}_s-\alpha^{\mathrm{plan}}_s\right|.
\end{equation}
This objective is invariant to global translation and rotation, matching the information represented by the plan. We additionally penalize insufficient spacing and implausibly large frame-to-frame root displacement:
\begin{align}
\mathcal{L}_{\mathrm{space}}&=\sum_{i\neq j,t}\left[d_{\min}-\left\|\mathbf{r}_{i,t}-\mathbf{r}_{j,t}\right\|_2\right]_{+},\\
\mathcal{L}_{\mathrm{jump}}&=\sum_{i,t}\left[\left\|\mathbf{r}_{i,t}-\mathbf{r}_{i,t-1}\right\|_2-v_{\max}\right]_{+}.
\end{align}

All objectives are masked to exclude padded dancers and invalid frames. The music-structure branch uses a separate segment self-similarity objective and is evaluated through its auxiliary ablation.

\subsection{Optimization and Sampling}

The Director, trainable music-conditioning modules, and Dance Generator are optimized jointly for 55 epochs using AdamW. We use a learning rate of $10^{-4}$, weight decay of $0.02$, and gradient clipping at $1.0$. Training is performed on two NVIDIA RTX~4090 GPUs with a per-GPU batch size of four and four-step gradient accumulation, giving an effective batch size of 32. Model selection and all hyperparameter choices are made on the validation set.

The diffusion process uses 1,000 training noise levels and 50 sampling steps. Conditioning dropout is set to $0.25$, and classifier-free guidance uses a weight of $2.0$. Unless otherwise stated, the Director predicts one choreography plan before diffusion sampling, and this plan remains fixed throughout the denoising trajectory. Optional refinement predicts a new plan only after a complete motion sample has been evaluated by the verifier. For the main results, refinement runs for up to $T$ cycles with $\tau_R$ selected on the validation set, and the highest-reward completed candidate is reported.

The complete objective is defined in Section~\ref{sec:objectives}. Loss terms for motion reconstruction, temporal velocity, forward-kinematic position, contact, formation realization, spacing, transition scheduling, and mobility are applied only to valid dancers and frames. The final loss weights are selected on the validation set and kept fixed for all test experiments.

\paragraph{Loss weights and initialization.}
Table~\ref{tab:loss_weights} reports the exact weights used by the main
\texttt{magd\_full5} model. The motion weights follow the reconstruction
configuration inherited from EDGE, except that the foot-contact weight is set
to $0.5$. Contact supervision is applied only on ground-truth support frames.
The music-structure branch is enabled for the main model.

\begin{table}[H]
    \centering
    \caption{\textbf{Loss weights for the main ChoreoStudio model.}}
    \label{tab:loss_weights}
    \footnotesize
    \setlength{\tabcolsep}{7pt}
    \renewcommand{\arraystretch}{1.18}
    \begin{tabular}{lclc}
        \hline
        Term & Weight & Term & Weight \\
        \hline
        $\mathcal{L}_{\mathrm{rec}}$   & 0.636
        & $\mathcal{L}_{\mathrm{vel}}$ & 2.964 \\
        $\mathcal{L}_{\mathrm{pos}}$   & 0.646
        & $\mathcal{L}_{\mathrm{contact}}$ & 0.5 \\
        $\mathcal{L}_{\mathrm{dir}}$   & 0.3
        & $\mathcal{L}_{\mathrm{real}}$ & 1.0 \\
        $\mathcal{L}_{\mathrm{space}}$ & 1.0
        & $\mathcal{L}_{\mathrm{jump}}$ & 1.0 \\
        $\mathcal{L}_{\mathrm{struct}}$ & 1.0
        & $\mathcal{L}_{\mathrm{struct\mbox{-}inv}}$ & 0.3 \\
        \hline
    \end{tabular}
\end{table}

Forward-kinematic velocity, acceleration, and feasibility losses are disabled
for the reported model. The perception fusion gate is initialized to $0.5$,
and the initial-state spread is set to $0.875$. All remaining settings use the
defaults reported in this appendix.

\subsection{Length-Matched Evaluation Protocol}

All methods are evaluated on the same 154 test identities. Main results use the
150-frame setting, and the separate long-sequence experiment uses 720 frames.
Within each setting, all methods share the same temporal support, motion
conversion, feature extraction, and metric implementations. Scores copied from
incompatible sequence lengths or preprocessing pipelines are not mixed with
our results, and each generated sequence is evaluated as a whole rather than
as independently scored short clips.

Sequence-level statistics are treated as the independent evaluation units. Pairwise and window-level measurements are first aggregated within each sequence and then averaged across test sequences, preventing longer sequences or larger groups from receiving disproportionate weight. Where uncertainty is reported, we use 95\% bootstrap confidence intervals over test sequences. Paired significance tests use matched test sequences and are reported only for completed comparisons.

\section{Additional Results}
\label{sec:additional_results}

\subsection{Controlled Plan Interventions}
\label{app:plan_interventions}

We isolate the causal plan-to-motion path using paired, single-cycle generation
on one 150-frame window. Each pair shares the music, sparse labels, initial
state, actor seeds, and diffusion noise; verifier replanning, best-of-cycle
selection, and window stitching are disabled. An edit is applied after plan
prediction, followed only by the deterministic Schedule and Encode operations
that reconstruct $\mathbf U$ and $\mathbf B$.

\paragraph{Formation geometry.}
We transplant a target shape from a training sequence with the same dancer
count, center it, and match its RMS centroid radius to the original plan before
replacing the selected segment. Distances, scale, and planned travel are then
recomputed, while relation, gate, mobility, role, and divergence variables are
fixed. For edited segments $\mathcal E$, realization error is
\begin{equation}
E_{\mathrm{form}}(\mathbf X,\mathbf P')
=
\frac{1}{|\mathcal E|N(N-1)}
\sum_{s\in\mathcal E}
\left\|
\widehat{\mathbf D}^{X}_{s}
-\widehat{\mathbf D}^{P'}_{s}
\right\|_1,
\end{equation}
which is invariant to global translation, rotation, and scale. Uniform scale
edits provide a dose--response check, while a $90^\circ$ rotation is an
invariance control.

\paragraph{Transition timing.}
We shift the selected gate event by an integer number of segments and recompute
the scheduled formation chain. The requested frame shift is
$\Delta t=\Delta s\,L_s$, where $L_s$ is obtained from the model's segment map.
For generated motion, we smooth the frame-wise change in scale-free pairwise
geometry and detect the dominant event within a local window around the
transition. For edited target time $\tau'$, the timing error is
\begin{equation}
E_{\mathrm{time}}(\mathbf X,\tau')
=
\frac{|\widehat{\tau}(\mathbf X)-\tau'|}{L}.
\end{equation}
Events with insufficient prominence or peaks at the search-window boundary are
marked invalid rather than treated as zero response.

\paragraph{Dancer mobility.}
We swap the complete mobility rows of the dancers with the highest and lowest
allocations, preserving the formation, gate, relation graph, and mobility
multiset. Let
$\bar{\mathbf r}_{i,t}=\mathbf r_{i,t}-N^{-1}\sum_j\mathbf r_{j,t}$ be the
centroid-relative root. Over transition interval $\mathcal T$, the realized
mobility share is
\begin{equation}
q_i(\mathbf X)
=
\sum_{t\in\mathcal T}
\|\bar{\mathbf r}_{i,t}-\bar{\mathbf r}_{i,t-1}\|_2,
\qquad
\widehat q_i(\mathbf X)
=
\frac{q_i(\mathbf X)}{\sum_jq_j(\mathbf X)+\epsilon},
\end{equation}
with error
\begin{equation}
E_{\mathrm{mob}}(\mathbf X,\boldsymbol\mu')
=\frac{1}{2}
\|\widehat{\mathbf q}(\mathbf X)-\boldsymbol\mu'\|_1.
\end{equation}

For $k\in\{\mathrm{form},\mathrm{time},\mathrm{mob}\}$, error reduction is
\begin{equation}
\mathrm{ER}_k
=
\frac{E_k(\mathbf X)-E_k(\mathbf X')}
{E_k(\mathbf X)+\epsilon},
\end{equation}
where both errors are measured against the edited target. Direction reports
the fraction of valid pairs changing as requested. Identical-plan no-op and
rotation-invariance controls must reproduce the baseline output within
numerical tolerance before a run is accepted. We additionally report paired
changes in TIF and Jitter to detect physical or temporal degradation.

As shown in Table~\ref{tab:intervention}, editing each plan component produces
the intended change in generated motion. Formation intervention yields the
strongest response, reducing realization error by 40\% and changing the output
in the requested direction for 78.54\% of the samples. Transition timing remains
similarly controllable, while mobility intervention produces a weaker but still
consistent response, reflecting the greater difficulty of reallocating movement
while preserving the formation target. Across all interventions, the small
changes in TIF and Jitter indicate that the requested behavior is realized
without materially degrading physical plausibility or temporal stability.
These paired results verify that the explicit plan controls what formation is
formed, when it changes, and which dancers execute the transition.

\subsection{Long-Sequence Generation}
\label{app:long_sequence}

We separately evaluate 720-frame generation using the same ChoreoStudio checkpoint as in the main experiments, without length-specific retraining. TCDiff++ is likewise evaluated from its existing checkpoint, whereas the remaining baselines are retrained for the target length. ChoreoStudio also uses one checkpoint across all evaluated group sizes.

As shown in Table~\ref{tab:long_results}, ChoreoStudio achieves the best performance across all metrics for one and two dancers. As the group size increases, it maintains the highest GMC and lowest TIF for both three and four dancers, although TCDiff++ obtains slightly lower GMR. This trade-off indicates that ChoreoStudio prioritizes stable cross-dancer coordination and collision control while retaining competitive motion realism. Overall, one checkpoint generalizes across sequence lengths and group sizes without length-specific or group-size-specific retraining.

\begin{table}[H]
\centering
\caption{\textbf{Dance generation performance in the 720-frame long-term
setting.} Best results are \textbf{bold}; second-best results are
\underline{underlined}.}
\label{tab:long_results}
\setlength{\tabcolsep}{4.2pt}
\renewcommand{\arraystretch}{1.10}
\resizebox{\textwidth}{!}{
\begin{tabular}{l|ccc|ccc|ccc|ccc}
\hline
\multirow{2}{*}{Method} &
\multicolumn{3}{c|}{\textbf{1 Dancer}} &
\multicolumn{3}{c|}{\textbf{2 Dancers}} &
\multicolumn{3}{c|}{\textbf{3 Dancers}} &
\multicolumn{3}{c}{\textbf{4 Dancers}} \\
\cline{2-13}
& FID$\downarrow$ & Div$\uparrow$ & PFC$\downarrow$
& GMR$\downarrow$ & GMC$\uparrow$ & TIF$\downarrow$
& GMR$\downarrow$ & GMC$\uparrow$ & TIF$\downarrow$
& GMR$\downarrow$ & GMC$\uparrow$ & TIF$\downarrow$ \\
\hline

GCD~\cite{le2023controllable}
& 51.25 & 7.24 & 3.52
& 38.43 & \underline{80.76} & 0.21
& 40.68 & 79.25 & 0.28
& 42.52 & 80.32 & 0.20 \\

CoDancers~\cite{yang2024codancers}
& 43.62 & 5.48 & 4.23
& 31.67 & 68.32 & \underline{0.12}
& 35.20 & 70.53 & \underline{0.14}
& 36.32 & 72.52 & \underline{0.13} \\

TCDiff++~\cite{dai2026tcdiffpp}
& \underline{22.47} & \underline{16.19} & \underline{1.53}
& \underline{16.52} & 79.63 & \underline{0.12}
& \textbf{14.67} & \underline{81.64} & 0.15
& \textbf{13.78} & \underline{80.43} & 0.14 \\

ST\mbox{-}GDance~\cite{xu2025stgdance}
& 31.42 & 14.35 & 3.21
& 32.56 & 75.32 & 0.14
& 34.25 & 74.42 & 0.15
& 33.61 & 76.26 & \underline{0.13} \\

\hline

ChoreoStudio
& \textbf{21.52} & \textbf{16.32} & \textbf{1.03}
& \textbf{14.48} & \textbf{83.42} & \textbf{0.11}
& \underline{15.63} & \textbf{82.64} & \textbf{0.12}
& \underline{14.58} & \textbf{81.37} & \textbf{0.12} \\

\hline
\end{tabular}
}
\end{table}

\subsection{Additional Planning Diagnostics}
\label{app:planning_diagnostics}

Table~\ref{tab:planning_diagnostics} examines whether the gains arise from explicit choreography planning rather than generic interaction modeling. The \emph{relation-only} variant removes formation targets, transition timing, and mobility allocation while retaining directed dancer relations. \emph{Latent bypass} replaces the structured plan-to-motion interface with unrestricted Director features. The remaining variants remove either relation-formation feedback or segment-level music structure.

All variants degrade group-motion realism and coordination. Relation-only reasoning therefore cannot replace explicit formation and scheduling decisions, while the latent-bypass result supports routing choreography through the structured interface. Removing relation-formation feedback causes the largest overall degradation, particularly in FMR, confirming the importance of mutually refining relational reasoning and realizable geometry. Removing music structure mainly reduces GMR and FMR, indicating that segment-level musical context is particularly important for macro-level formation evolution.

\begin{table}[H]
    \centering
    \caption{\textbf{Additional planning diagnostics.} Results are evaluated
    using the same 150-frame three-dancer protocol as the main experiments.
    FMR is interpreted by proximity to the real-data reference.}
    \label{tab:planning_diagnostics}
    \footnotesize
    \setlength{\tabcolsep}{4pt}
    \renewcommand{\arraystretch}{1.40}
    \begin{tabular}{lcccc}
        \hline
        Variant
        & GMR$\downarrow$
        & GMC$\uparrow$
        & TIF$\downarrow$
        & FMR \\
        \hline
        Full ChoreoStudio
        & 10.42 & 83.23 & 0.08 & 0.848 \\
        \hline
        Relation-only
        & 15.29 & 81.89 & 0.10 & 0.792 \\
        Latent bypass
        & 14.52 & 80.45 & 0.09 & 0.819 \\
        w/o relation-formation feedback
        & 15.73 & 79.79 & 0.11 & 0.738 \\
        w/o music structure
        & 14.33 & 82.71 & 0.09 & 0.798 \\
        \hline
    \end{tabular}
\end{table}

\subsection{Denoising-Budget-Matched Refinement Analysis}
\label{app:refinement_robustness}

We examine whether the gains from verification-guided refinement arise merely
from additional diffusion computation. We compare two 50-step refinement cycles with a single 100-step generation, keeping the total number of denoising steps fixed. We additionally include best-of-two independent 50-step samples selected by the same frozen Verifier, controlling for the benefit of generating and ranking multiple candidates. Improvement over both controls therefore provides evidence for feedback-guided plan revision rather than additional denoising or sample selection.
Unlike the adaptive best-over-cycles protocol used for the main results, the
$R=2$ setting here always executes exactly two cycles to enforce a fixed
denoising budget.

\begin{table}[H]
    \centering
    \caption{\textbf{Denoising-budget-matched refinement analysis.} NFE denotes the total number of denoising evaluations. Best-of-two samples are independently generated and ranked by the same frozen Verifier. FMR is ranked by proximity to the real-data reference of 0.832.}
    \label{tab:refinement_robustness}
    \small
    \setlength{\tabcolsep}{3.5pt}
    \renewcommand{\arraystretch}{1.15}
    \begin{tabular}{lccccc}
        \hline
        Setting
        & NFE
        & GMR$\downarrow$
        & GMC$\uparrow$
        & TIF$\downarrow$
        & FMR \\
        \hline
        $R=1$, 50 steps
        & 50
        & 15.43
        & 79.32
        & \underline{0.10}
        & 0.810 \\

        $R=1$, 100 steps
        & 100
        & 14.74
        & 80.94
        & \textbf{0.09}
        & 0.805 \\

        Best-of-2, 50 steps
        & 100
        & \underline{13.62}
        & \underline{81.29}
        & \textbf{0.09}
        & \textbf{0.826} \\

        $R=2$, 50 steps
        & 100
        & \textbf{10.36}
        & \textbf{83.20}
        & \underline{0.10}
        & \underline{0.847} \\
        \hline
    \end{tabular}
\end{table}

Increasing a single-cycle sampler from 50 to 100 steps yields only modest
improvements, while selecting the better of two independent samples provides a
stronger gain. Under the same 100-step denoising budget, feedback-guided
refinement further reduces GMR by 23.9\% and improves GMC by 1.91 points
over best-of-two sampling. Its FMR remains close to the real-data reference,
although TIF increases slightly from 0.09 to 0.10. These results indicate
that the improvement in group realism and coordination cannot be explained by
additional denoising or candidate selection alone.

\subsection{Semantic Control}
\label{app:semantic_control}

We evaluate whether the high-level description provides controllable
macro-intent without treating it as a source of frame-accurate choreography.
Each paired comparison fixes the music, sampled actor seeds, diffusion noise,
and sampling configuration while changing only the description. We measure the
resulting changes in group scale, transition rate, pose divergence, and unison.
Aggregate GMR and TIF changes test whether the requested behavior is achieved by
degrading motion realism or increasing collisions.

For each paired edit, direction accuracy is the percentage of examples in
which the corresponding target metric changes in the requested direction.
Changes in GMR and TIF are reported to detect semantic control achieved at the
expense of motion quality or physical plausibility.

\begin{table}[H]
    \centering
    \caption{\textbf{Semantic control under paired generation.} Changes are
    computed as edited minus original. Direction accuracy reports the percentage
    of pairs whose target metric changes in the requested direction.}
    \label{tab:semantic_control}
    \small
    \setlength{\tabcolsep}{4pt}
    \renewcommand{\arraystretch}{1.15}
    \begin{tabular}{lllccc}
        \hline
        Description edit
        & Target metric
        & Target change
        & Direction$\uparrow$
        & $\Delta$GMR
        & $\Delta$TIF \\
        \hline
        Compact to expansive
        & Spread
        & +0.34\,m
        & 82.6\%
        & +0.28
        & -0.006 \\
        
        Stable to dynamic
        & IDDV
        & +0.052
        & 77.4\%
        & +0.46
        & +0.004 \\
        \hline
    \end{tabular}
\end{table}

Table~\ref{tab:semantic_control} shows that contrasting descriptions produce
consistent changes in the intended group behavior. Expansive descriptions
increase group spread by 0.34\,m in 82.6\% of paired samples, while dynamic
descriptions increase IDDV by 0.052 with a direction accuracy of 77.4\%.
The small changes in GMR and TIF indicate that these semantic responses are
achieved without substantial degradation in motion realism or physical
plausibility. These results support the semantic description as a controllable
macro-level planning condition.

\subsection{User Study}
\label{app:user_study}

We complement the automatic evaluation with a blind user study following the
ranking protocol of TCDiff~\citep{dai2025harmonious}. We recruit 30 participants
aged 18 to 55. In each trial, participants view four anonymized and randomly
ordered videos generated by CoDancers, TCDiff++, ST-GDance, and ChoreoStudio
under the same music and group setting. They rank the videos without ties along
four dimensions: \emph{Motion Realism}, \emph{Music Alignment},
\emph{Formation Quality}, and \emph{Inter-Dancer Coordination}. Formation
Quality evaluates the visual organization and evolution of group formations,
while Inter-Dancer Coordination assesses whether dancers interact coherently
during formation transitions.

Ranks are assigned scores from 4 (best) to 1 (worst). For method $m$ and
dimension $d$, we normalize the scores to $[0,1]$ as
\begin{equation}
S_{m,d}
=
\frac{1}{Q}
\sum_{q=1}^{Q}
\frac{r_{q,m,d}-1}{3},
\end{equation}
where $Q$ is the number of valid rankings and $r_{q,m,d}$ is the rank score in
trial $q$.

\begin{figure}[H]
    \centering
    \includegraphics[width=\linewidth]{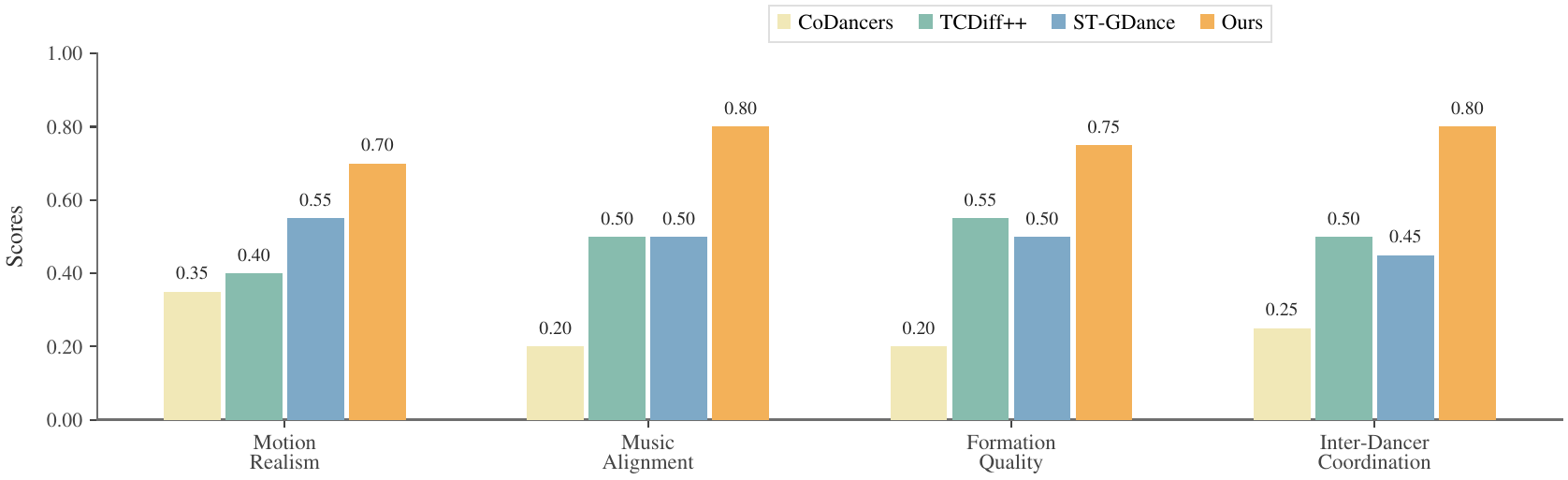}
    \caption{\textbf{User study results.} Normalized blind-ranking scores across
    four perceptual dimensions. ChoreoStudio is consistently preferred, with
    the largest margins in Music Alignment and Inter-Dancer Coordination.}
    \label{fig:user_study}
\end{figure}

As shown in Figure~\ref{fig:user_study}, ChoreoStudio receives the highest score
on all four dimensions. It achieves clear gains in Motion Realism and Formation
Quality, indicating that explicit planning improves group organization without
compromising motion naturalness. The largest margins occur in Music Alignment
and Inter-Dancer Coordination, supporting the roles of multi-scale music
conditioning and coupled formation, timing, and mobility planning.

\section{More Qualitative Results}
\label{app:qualitative_results}

\subsection{Qualitative Results and Video Demonstrations}

\textbf{Visualization and rendering.}
Unless otherwise specified, the paper figures use the standard SMPL mesh to
convert generated PKL motion sequences into FBX animations under a shared
camera and rendering configuration. For the supplementary demonstrations in
\texttt{index.html}, we additionally retarget the generated motions to the YBot
character for higher-quality visualization. The videos are rendered in Blender
using scripts adapted from the open-source TCDiff implementation
~\citep{dai2025harmonious}. Rendering affects only the visual appearance and
does not modify the generated root trajectories or body motions.

\begin{figure}[H]
    \centering
    \begin{subfigure}[c]{0.4\linewidth}
        \centering
        \includegraphics[width=\linewidth]{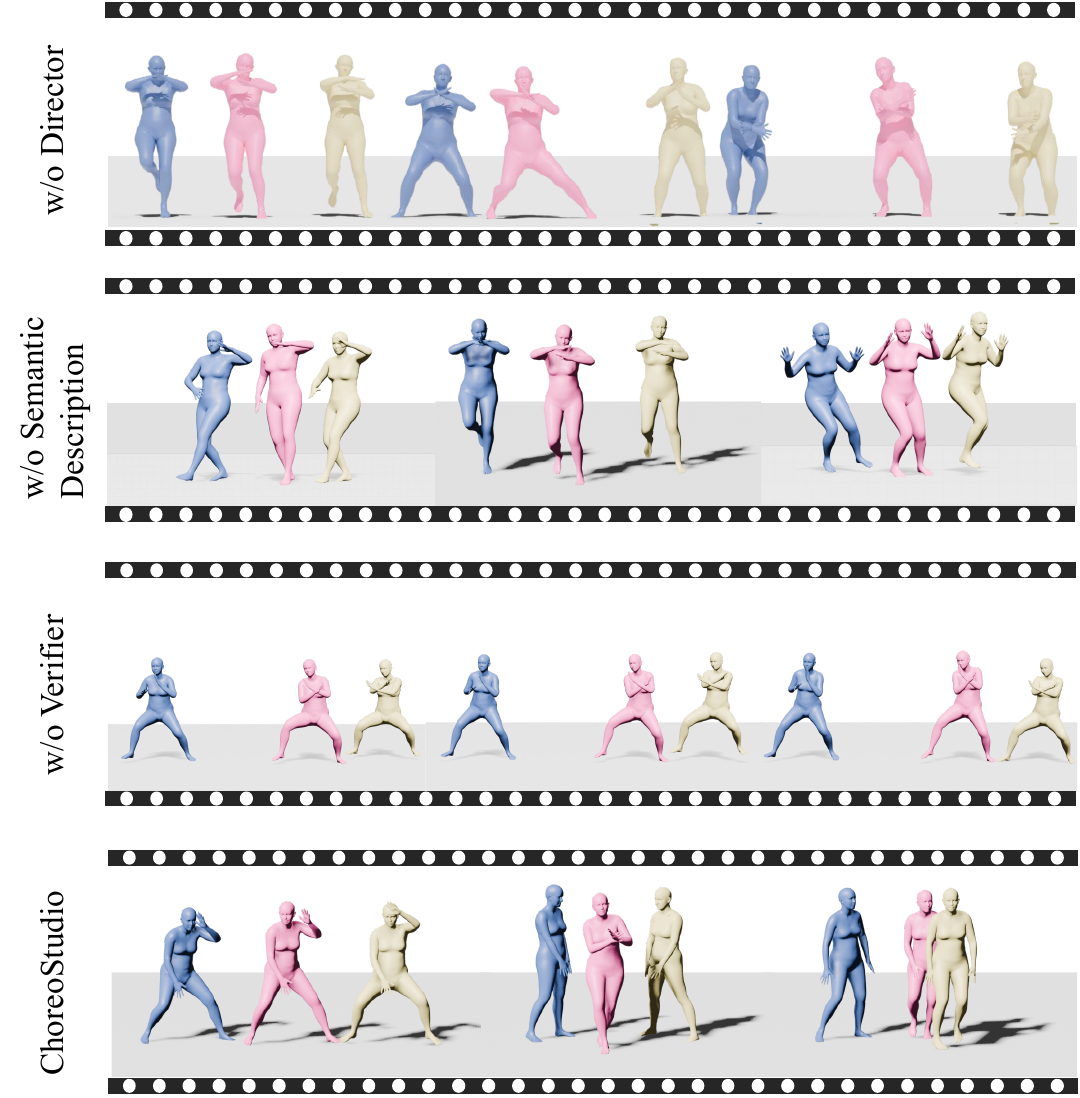}
        \caption{Ablation study}
        \label{fig:qualitative_ablation}
    \end{subfigure}
    \hfill
    \begin{subfigure}[c]{0.55\linewidth}
        \centering
        \includegraphics[width=\linewidth]{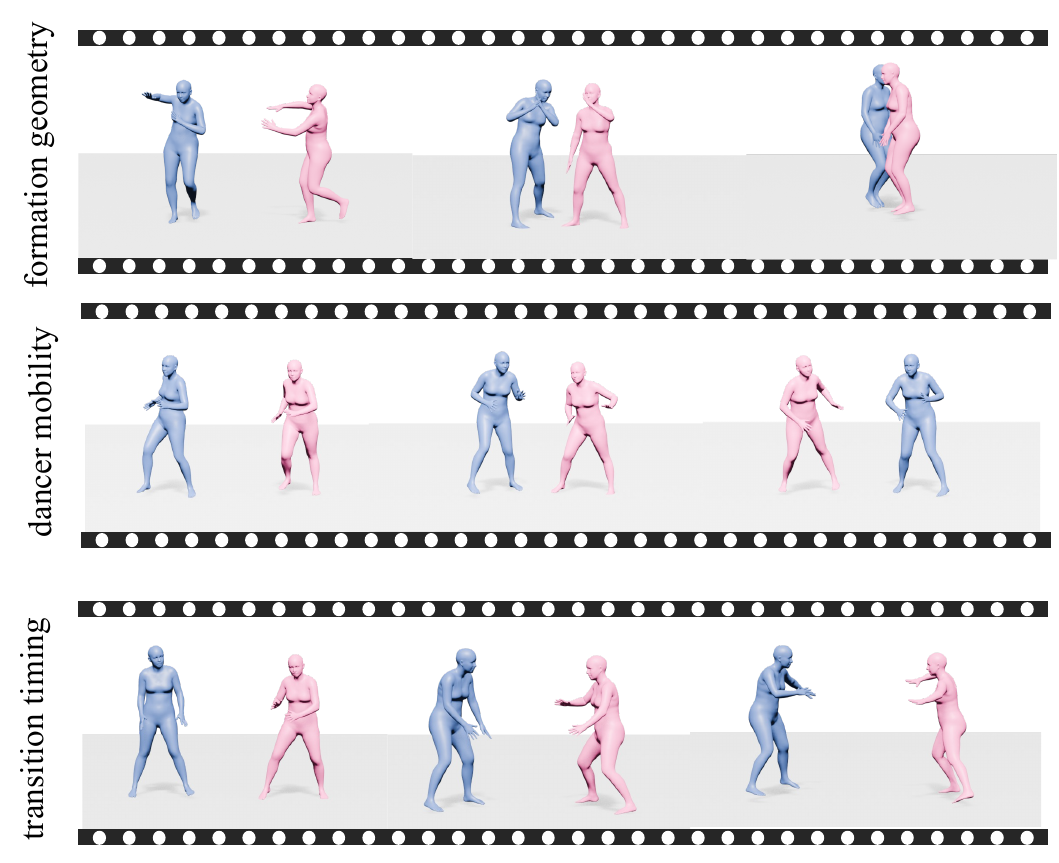}
        \caption{Controlled plan interventions}
        \label{fig:interaction_control}
    \end{subfigure}
    \caption{\textbf{Qualitative analysis of explicit choreography planning.}
    (a) Removing the Director, semantic description, or Verifier produces
    distinct degradations in group organization, motion variation, and
    formation transitions.
    (b) Editing formation geometry, transition timing, or dancer mobility
    produces the requested motion changes while fixing the music, initial
    noise, and remaining plan variables.}
    \label{fig:qualitative_analysis}
\end{figure}

\textbf{Qualitative ablations.}
Figure~\ref{fig:qualitative_ablation} compares the effects of removing the
Director, semantic description, and Verifier. Without the Director, dancers
perform mismatched poses and undergo inconsistent spatial changes, producing a
sequence that remains locally plausible but lacks coherent group-level
organization. Removing the semantic description makes the dancers follow more
uniform and repetitive movement patterns, reducing dancer-specific variation
and weakening the progression of the choreography. Without the Verifier, the
sequence remains close to the same wide formation and repeated pose across
frames, suggesting that implausible or stagnant plans are not effectively
corrected. In contrast, ChoreoStudio maintains distinct but coordinated dancer
motions and gradually reorganizes the group from a dispersed arrangement into
a compact formation. The comparison complements the quantitative ablations by
showing how explicit planning provides group structure, semantic context
encourages differentiated movement, and verification improves the realization
of formation transitions.

\textbf{Plan interventions and group-size generalization.}
Figure~\ref{fig:interaction_control} visualizes the three forms of plan control.
The formation-geometry intervention changes the relative spatial target,
causing the two dancers to evolve from a separated arrangement toward a compact
formation. Editing dancer mobility changes how the transition is distributed
between the dancers, producing a different pattern of spatial participation
and position exchange while preserving the overall group task. The
transition-timing intervention changes the temporal location of the spatial
reorganization, allowing the same formation change to occur earlier or later
in the sequence. Because the music, initial noise, and non-intervened plan
variables are fixed, these differences can be attributed to the corresponding
plan edits rather than sampling variation. The resulting motions remain
temporally continuous, showing that the plan-to-motion interface can modify
\emph{what} changes, \emph{when} it changes, and \emph{who} executes the
transition without regenerating an unrelated dance.

%\textbf{Group-size generalization.}
%Figure~\ref{fig:group_size} demonstrates that a single ChoreoStudio checkpoint generalizes across different dancer counts without size-specific retraining. As the group size increases, the number of pairwise interactions and possible spatial conflicts grows substantially. Nevertheless, the model preserves coherent individual motion, meaningful relative spacing, and coordinated group-level organization. The dancers remain distinguishable and contribute differently to the collective movement rather than collapsing into identical motion patterns. These results indicate that the relative formation representation, dancer-specific planning, and plan-to-motion interface extend naturally to variable-size groups.

%\begin{sidewaysfigure*}
    %\centering
    %\includegraphics[width=\textheight]{figs/group_size_results.pdf}
    %\caption{Qualitative results of ChoreoStudio under different group sizes.}
    %\label{fig:group_size}
%\end{sidewaysfigure*}

\end{document}